%% file: main.tex
\documentclass[times,twocolumn,final, nopreprintline]{elsarticle}

\usepackage{framed,multirow}

\usepackage{amssymb}
\usepackage{latexsym}
\usepackage{algorithm}
\usepackage[noend]{algpseudocode} 

\usepackage{color}
\usepackage{comment}

\usepackage{graphicx}

\usepackage{amsmath}
\usepackage{caption}
\usepackage{subcaption}
\usepackage[export]{adjustbox}
\usepackage{tabularx} 

\usepackage{url}
\usepackage{xcolor}

\definecolor{newcolor}{rgb}{.8,.349,.1}

\begin{document}

\begin{frontmatter}
\title{HookNet: multi-resolution convolutional neural networks for semantic segmentation in histopathology whole-slide images}%

\author[1]{Mart van Rijthoven \corref{cor1}}
\cortext[cor1]{Corresponding author: Mart van Rijthoven \\
               Email: mart.vanrijthoven@radboudumc.nl}
\author[1]{Maschenka Balkenhol} 
\author[2]{Karina Siliņa}
\author[1,3]{Jeroen van der Laak}
\author[1]{Francesco Ciompi}

\address[1]{Radboud University Medical Center, Geert Grooteplein 21 , Nijmegen 6525EZ, Netherlands}
\address[2]{Institute of Experimental Immunology, University of Zurich, Zurich, Switzerland}
\address[3]{Center for Medical Image Science and Visualization, Linköping University, Linköping, Sweden}



\begin{abstract}

We propose HookNet, a semantic segmentation model for histopathology whole-slide images, which combines \emph{context} and \emph{details} via multiple branches of encoder-decoder convolutional neural networks.
Concentric patches at multiple resolutions with different fields of view are used to feed different branches of HookNet, and intermediate representations are combined via a \emph{hooking} mechanism.

We describe a framework to design and train HookNet for achieving high-resolution semantic segmentation and introduce constraints to guarantee pixel-wise alignment in feature maps during hooking. We show the advantages of using HookNet in two histopathology image segmentation tasks where tissue type prediction accuracy strongly depends on contextual information, namely (1) multi-class tissue segmentation in breast cancer and, (2) segmentation of tertiary lymphoid structures and germinal centers in lung cancer. We show the superiority of HookNet when compared with single-resolution U-Net models working at different resolutions as well as with a recently published multi-resolution model for histopathology image segmentation.

\end{abstract}


\end{frontmatter}


\begin{figure*}
\begin{center}
\centering

\begin{subfigure}{.16\linewidth}
    \includegraphics[width=\linewidth]{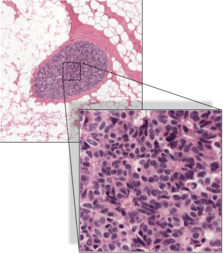}
    \caption{DCIS}
\end{subfigure}
\begin{subfigure}{.16\linewidth}
    \includegraphics[width=\linewidth]{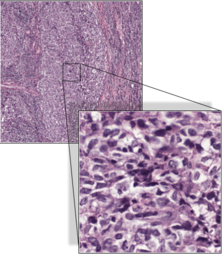}
    \caption{IDC}
\end{subfigure}
\begin{subfigure}{.16\linewidth}
    \includegraphics[width=\linewidth]{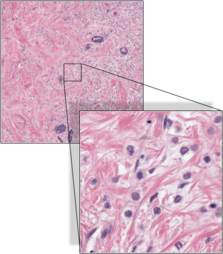}
    \caption{ILC}
\end{subfigure} \qquad \qquad
\begin{subfigure}{.16\linewidth}
    \includegraphics[width=\linewidth]{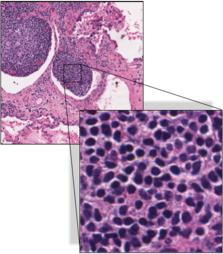}
    \caption{TLS}
\end{subfigure}
\begin{subfigure}{.16\linewidth}
    \includegraphics[width=\linewidth]{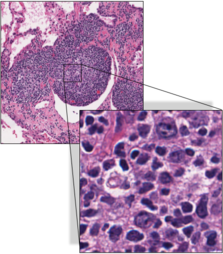}
    \caption{GC}
\end{subfigure}
\caption{Examples of ductal carcinoma in situ (DCIS), invasive ductal carcinoma (IDC) and invasive lobular carcinoma (ILC) in breast tissue, and tertiary lymphoid structures (TLS) and germinal centers (GC) in lung tissue. For each example, multi-resolution/multi-field-of-view (MFMR) patches (introduced in section \ref{section_mfmr}) are shown: both a low resolution/large-field-of-view and a concentric, high-resolution/small-field-of-view patch are depicted.}
\label{fig:different_tissue_types}
\end{center}

\end{figure*}

\section{Introduction}

Semantic image segmentation is the separation of concepts by grouping pixels belonging to the same concept, with the aim of simplifying image representation and understanding. In medical imaging, tumor detection and segmentation are necessary steps for diagnosis and disease characterization. This is especially relevant in histopathology, where tissue samples with a wide variety and amount of cells within a specific context have to be analyzed by pathologists for diagnostic purposes. 

Introduction of high-resolution and high throughput digital scanners have de-facto revolutionized the field of pathology by digitizing tissue samples and producing gigapixel whole-slide images (WSI). In this context, the digital nature of WSIs allows for the possibility to use computer algorithms for automated histopathology image segmentation, which can be a valuable diagnostic tool for pathologists to identify and characterize different types of tissue, including cancer.


\subsection{Context and details in histopathology}
It has long been known that despite individual cancer cells may share morphological characteristics, the way they grow into specific patterns make a profound difference in the prognosis of the patient. As an example, in hematoxylin and eosin (H\&E) stained breast tissue samples, different histological types of breast cancer can be distinguished. For instance, an invasive tumor that originates in the breast duct (invasive ductal carcinoma, IDC) can show a wide variety in growth patterns. In contrast, an invasive tumor that originates in the breast lobules (invasive lobular carcinoma, ILC), is characterized by individually arranged tumor cells. Furthermore, the same type of ductal carcinoma cells can be confined within the breast duct (ductal carcinoma in situ, DCIS) or become invasive by spreading outside the duct (IDC) (see Figure \ref{fig:different_tissue_types}) (\cite{lakhani_who_2012}). To differentiate between these types of cancer, pathologists typically combine observations. For example, they look at the global architectural composition of the tissue sample and analyze the \emph{context} of each tissue component, including cancer, to identify the presence of duct (both healthy and potentially cancerous) and other tissue structures. Additionally, they zoom-in into each region of interest, where the tissue is examined at a high-resolution, to obtain the \emph{details} of the cancer cells, and characterize the tumor based on its local cellular composition.
Another example where pathologists take advantage of both context and details is the spatial distribution of immune cells, which may be detected in the presence of inflammation inside the tumor or the stromal compartment of the cancer regions, as well as in specific clustered groups called tertiary lymphoid structures (TLS), which may develop in response to cancer. In subsequent stages of TLS maturation, germinal centers (GC) are formed within the TLS (see Figure \ref{fig:different_tissue_types}). It has been shown that the development of GC-containing TLS have a significant relevance for patient survival and is an essential factor for the understanding of tumor development and treatment (\cite{sautes-fridman_tertiary_2016}, \cite{silina_germinal_2018}). A GC always lays within a TLS. TLS contain a high density of lymphocytes with poorly visible cytoplasm, while GCs rather share similarities with other less dense tissues like tumor nests. To identify the TLS region and to differentiate between TLS and GC, both fine-grained details, as well as contextual information, are needed.



\subsection{The receptive field and the field of view}

In recent years, the vast majority of state of the art image analysis algorithms are based on convolutional neural networks (CNN), a deep learning model that can tackle several computer vision tasks, including semantic segmentation (Long et al. (2015), Jégou et al. (2017), Chen et al.(2018)). In semantic segmentation, label prediction at pixel level depends on the receptive field, which is the extent of the area of the input that is observable by a model. The size of the receptive field of a CNN depends on the filter size, the pooling factor, and the number of convolutional and pooling layers. {\color{black} By increasing these parameters, the receptive field also increases, allowing the model to capture more contextual information.
However, this often comes at the cost of an increase in the input size, which  causes a high memory consumption due to large feature maps. As a consequence, a number of implicit restrictions in model optimization have to be applied often, such as reduction of the number of model's parameters, number of feature maps, mini-batch size, or size of predicted output, which may result in an ineffective training and in an inefficient inference.}

Another aspect concerning the observable information is the field of view (FoV), which is the distance over the area (i.e., the actual space that the pixels disclose) in an input image and depends on the spatial resolution of the input image. The FoV has implications for the receptive field: the same model, {\color{black} the same input size,} and the same receptive field, can comprise a wider FoV by considering an image at lower resolution due to the compressed distance over the area (i.e., fewer pixels disclose the same FoV of the original resolution). Thereby, using a down-sampled representation of the original input image, the model can benefit from more contextual aggregation (\cite{graham_sams-net:_2018}), at the cost of losing high-resolution details. Furthermore, contextual aggregation is limited by the input dimensions, meaning that a receptive field size can only exceed the input dimensions if padded artificial input pixels are used {\color{black}(a technique usually referred to as the use of \emph{same} padding)}, which do not contain contextual information. While reducing the original input dimensions can be used to focus on scale information (\cite{kausar_multi-scale_2018}, \cite{li_multi-scale_2018}), the potential contextual information remains unchanged.

\subsection{Multi-field-of-view multi-resolution patches}\label{section_mfmr}
Whole-slide images (WSI) are pyramidal data structures containing multi-resolution gigapixel images, including down-sampled representations of the original image.
In the context of CNN models development, it is not possible to capture a complete WSI at full resolution in the receptive field, due to the billions of pixels in a WSI, which exceeds the capacity of the memory of a single modern GPU that is usually used to train CNN models. A common way to overcome this limitation is to train a CNN with patches (i.e., sub-regions from a WSI). 
Due to the multi-resolution nature of WSIs, the patches can originate from different spatial resolutions, which are expressed in micrometers per pixel ($\mu m/px$). A patch is extracted by selecting a position within the WSI, together with a size and a particular resolution. When extracting a patch at the highest available resolution, the potential contextual information is not yet depleted because there is available tissue around the patch that is not considered in the receptive field. By extracting a concentric (i.e., centered on the same location of the whole-slide image) patch, with the same size but lower resolution, the same receptive field aggregates more contextual information and includes information that was not available before. Multiple concentric same sized patches, extracted at different resolutions, can be interpreted as having multiple FoVs. Hence we call a set of these  patches: multi-field-view multi-resolution (MFMR) patches (see Fig \ref{fig:different_tissue_types}).

To date, research using MFMR patches extracted from histopathology images, has mostly been focused on combining features obtained from MFMR patches for patch \emph{classification} (\cite{alsubaie_multi-resolution_2018}, \cite{wetteland_multiscale_2019}, \cite{sirinukunwattana_improving_2018}). However, using patch classification for the purpose of semantic segmentation results in a coarse segmentation map or heavy computation due to a required sliding window approach, which is needed for the segmentation of every pixel. For the task of segmentation, the use of MFRM patches is not straightforward: when combining features obtained from MFMR patches, pixel-wise alignment should be enforced when integrating them. \cite{gu_multi-resolution_2018} proposed to use a U-Net architecture (\cite{ronneberger_u-net:_2015}) for processing a high-resolution patch, and additional encoders to process lower resolution patches. Subsequently, feature maps from the additional encoders are cropped, up-sampled, and concatenated in the decoder parts of U-Net, at the places where skip connections are concatenated as well. The feature maps from the additional encoder are up-sampled  without skip connections, at the cost of localization precision. Furthermore, their proposed model concatenates feature maps at every depth in the decoder, which might be redundant and results in a high memory consuming model. Moreover, considering the necessity of pixel-wise alignment, their model is restricted to \emph{same} padding, which can introduce artifacts.

A multi-class problem where classes are known to be subjected to context and fine-grained details can benefit from the combined information in a set of MFMR patches. However, this is still an unsolved problem. The challenge is to simultaneously output a high-resolution segmentation based on fine details detectable at high resolution and the incorporation of unconcealed contextual features.

\subsection{Our contribution}

In this paper, we introduce a multi-branch segmentation framework based on convolutional neural networks that can simultaneously incorporate contextual information and high-resolution details to produce fine-grained segmentation maps of histopathology images.
The model, which we call \emph{HookNet}, consists of encoder-decoder branches, that allow for multi-resolution representations at different depths, which we take advantage of by concatenating relevant features between branches in the decoder parts via a \emph{hooking} mechanism. 
We give a detailed description of how to design and use the framework. In particular, we will show how to instantiate two U-Net branches and, in the design, limit the implementation to what is possible using a single modern GPU. Furthermore, we show the performance on multi-class and multi-organ problems, including tissue subjected to high-resolution details as well as context.

\begin{figure}
\includegraphics[width=\linewidth]{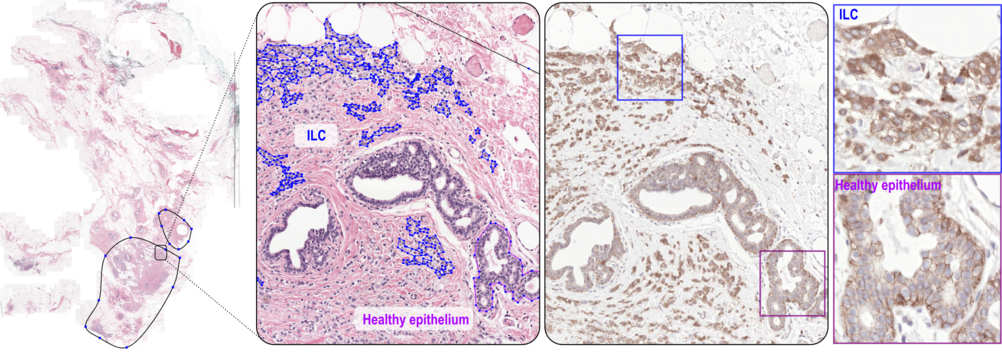}
\caption{Example of procedure to manually annotate ILC regions in breast cancer slides. Left: Slide stained with H\&E, with manual annotation of the region containing the bulk of tumor and  details of manual (sparse) annotations of ILC and of healthy epithelium. Right: Immunohistochemistry of the same tissue sample, de-stained from H\&E and re-stained with P120, used by {\color{black}medical research assistant} to guide manual annotations and to identify ILC cells, and details of the effects of P120 for ILC and healthy epithelium.}
\label{fig:ilc_examples}
\end{figure}

\begin{figure*}[h]
\centering
\includegraphics[width=0.7\linewidth]{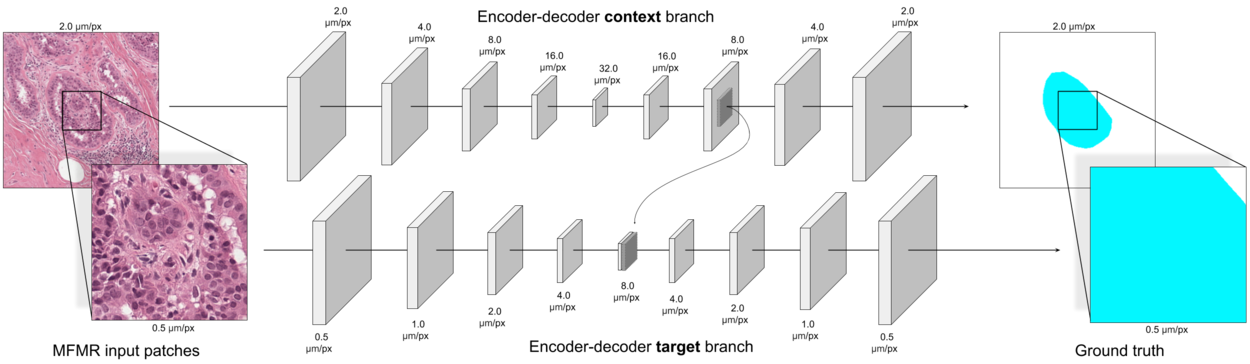}
\caption{HookNet model architecture. Concentric patches with multiple views at multiple resolutions (MFMR patches) are used as input to a dual encoder-decoder model. Skip connections for both branches are omitted for clarity. Feature maps are down- and up-sampled by a factor 2. In this example, the feature maps at depth 2 in the decoder part of the context branch comprise the same resolution as the feature maps in the bottleneck of the target branch. To combine contextual information with high-resolution information, feature maps from the context branch are \emph{hooked} in the target branch before the first decoder layer by cropping and concatenation.}
\label{fig:hooknet_model}
\end{figure*}

\section{Materials}
In order to train and assess the performance of HookNet, we collected data for two applications in histopathology image segmentation, namely multi-class tissue segmentation in breast cancer sections, and segmentation of TLS and GC in lung cancer sections.



\subsection{Breast dataset}
We collected 86 breast cancer tissue sections containing IDC (n=34), DCIS (n=35) and ILC (n=17). For the DCIS and IDC cases, we used H\&E stained tissue sections, which were initially made for routine diagnostics. All tissue sections were prepared according to the laboratory protocols from the Department of Pathology of Radboud University Medical Center, Nijmegen (the Netherlands). Slides were digitized using a Pannoramic P250 Flash II scanner (3DHistech, Hungary) at a spatial resolution of 0.24 $\mu m/px$. For ILC, new tissue sections were cut and stained for H\&E, after which, slides were scanned using the same scanner and scanning protocol as for the IDC/DCIS cases. After inspection of the WSIs, the H\&E stained ILC sections were de-stained and subsequently re-stained using P120 catenin antibody (P120) (\cite{canasmarques_e-cadherin_2016}), which stains lobular carcinoma cells cytoplasmic, rather than a membranous staining pattern in normal epithelial cells. P120 stained sections were subsequently scanned using the same scanner and protocol as the H\&E sections (\cite{brand_sequential_2014}). This procedure allowed us to have both H\&E and immunohistochemistry (IHC) of the same tissue section. {\color{black} Three people were involved in the creation of manual annotations: two medical research assistants, who had undergone a supervised training procedure in the pathology department to specifically recognize and annotate breast cancer tissue in histopathology slides, and a resident pathologist (MB), with six years of experience in diagnostics and research in digital pathology.} To guide the procedure of annotating ILC, {\color{black} the} resident pathologist visually identified and annotated the region containing the bulk of tumor in the HE slide. Successively, {\color{black} the} research assistants used this information next to the available IHC slide to identify and annotate ILC cells. Additionally, {\color{black} the research assistants} made annotations of DCIS, IDC, fatty tissue, benign epithelium, and an additional class of other tissue, containing inflammatory cells, skin/nipple, erythrocytes, and stroma. {\color{black} All annotations were finally checked by the resident pathologist and corrections were made when needed}. The in-house developed open-source software ASAP (\cite{litjens_1399_2018}) was used to make manual annotations. As a result, {\color{black}6279} regions were annotated, of which 1810 contained ILC cells. Sparse annotations of tissue regions were made, meaning that drawn contours could be both non-exhaustive (i.e., not all instances of that tissue type were annotated) and non-dense (i.e., not all pixels belonging to the same instance were included in the drawn contour). Examples of sparse annotations are depicted in Figure \ref{fig:ilc_examples}. As a result 6 classes were annotated in this dataset. For training, validation, and testing purposes, the WSIs were divided into training (n=50), validation (n=18), and test (n=18) sets, all containing a similar distribution of cancer types.




\subsection{Lung dataset}
We randomly selected 27 diagnostic H\&E-stained digital slides from the cancer genome atlas lung squamous cell carcinoma (TCGA-LUSC) data collection, which is publicly available in genomic data commons (GDC) Data Portal (\cite{grossman_toward_2016}). For this dataset, sparse annotations of TLS, GC, tumor, and other lung parenchyma were
made by a {\color{black}senior} researcher {\color{black} (KS)} with {\color{black} more than six years of} experience in tumor immunology and histopathology, and checked by a resident pathologist {\color{black}(MB)}. As a result, {\color{black} 1.098 annotations}, including 4 classes were annotated in this dataset. {\color{black}For model development and performance assessment, we used 3-fold cross-validation, which allowed us to test the performance of the presented models on all available slides. All three folds contain 12:6:9 images for training:validating:testing. We made sure that all splits had an appropriate class balance.}

\section{HookNet: multi-branch encoder-decoder network}
In this section we present ``HookNet'', a convolutional neural network model for semantic segmentation that processes concentric MFMR patches via multiple \emph{branches} of \emph{encoder-decoder} models and combines information from different branches via a ``hooking'' mechanism (see Figure \ref{fig:hooknet_model}).
The aim of HookNet is to produce semantic segmentation by combining information from (1) low-resolution patches with a large field of view, which carry contextual visual information, and (2) high-resolution patches with a small field of view, which carry fine-grained visual information.
For this purpose, we propose HookNet as a model that consists of two encoder-decoder branches, namely a \emph{context} branch, which extracts features from input patches containing contextual information, and a \emph{target} branch, which extracts fine-grained details from the highest resolution input patches for the target segmentation.
The key idea of this model is that fine-grained and contextual information can be combined by concatenating feature maps across branches{\color{black}, thereby resembling the process pathologist go through when zooming in and out while examining tissue.} 

We present the four main HookNet components following the order in which they should be designed to fulfill the constraints necessary for a seamless and accurate segmentation output, namely (1) branches architecture and properties, (2) the extraction of MFMR patches, (3) constraints of the ``hooking'' mechanism, and (4) the handling of targets and losses.

\subsection{Context and target branches} \label{section:branches}
The first step in the design of HookNet is the definition of its branches. Without loss of generality, we designed the model under the assumptions that (1) the two branches have the same architecture but do not share their weights, and (2) each branch consists of an encoder-decoder model based on the U-Net (\cite{ronneberger_u-net:_2015}) architecture. As in the original U-Net model, each convolutional layer performs valid 3x3 convolutions with stride 1, followed by max-pooling layers with a 2x2 down-sampling factor. For the up-sampling path, we adopted the approach proposed in \cite{odena_deconvolution_2016} consisting of nearest-neighbour 2x2 up-scaling followed by convolutional layers. 

\subsection{MFMR input patches}
The input to HookNet is a pair ($P_C, P_T$) of ($M\times M \times 3$) MFMR RGB concentric patches extracted at two different spatial resolutions $r_C$ and $r_T$ measured in $\mu m/px$ for the context (C) and the target (T) branch, respectively.
In this way, we ensure that the field of view of $P_T$ corresponds to the central square region of size $(M\frac{r_T}{r_C} \times M\frac{r_T}{r_C} \times 3)$ of $P_C$ but at lower resolution.
In order to get a seamless segmentation output and to avoid artifacts due to misalignment of feature maps in the encoder-decoder branches, specific design choices should be made on (1) the size and (2) the resolution of the input patches.
First, $M$ has to be chosen, such that all feature maps in the encoder path have an \emph{even} size before each pooling layer.
Stated initially in \cite{ronneberger_u-net:_2015}, this constraint is crucial for HookNet, as an unevenly sized feature map will also cause misalignment of feature maps not only via skip connections but also across branches.
Hence, this constraint ensures that feature maps across the two branches remain pixel-wise aligned.
Second, $r_T$ and $r_C$ should be chosen in such a way that given the branches architecture, a pair of feature maps in the decoding paths across branches comprise the same resolution.
This pair is an essential requisite for the ``hooking'' mechanism detailed in section \ref{section:hooking}.
In practice, given the depth $D$ (i.e., the number of pooling layers) of the encoder-decoder architecture, $r_C$ and $r_T$ should take on values such that the following inequality is true: $2^Dr_T \geq r_C$.

\subsection{Hooking mechanism} \label{section:hooking}
We propose to combine, i.e., hook-up information from the context branch into the target branch via the simple concatenation of feature maps extracted from the decoding paths of the two branches. {\color{black}Our chose for concatenation as the operation to combine feature maps is based on the success of skip connections in the original U-Net, which are also using concatenation. Moreover, concatenation allows downstream layers to operate over all feature maps, and therefore learn the optimal operation to apply during the parameters optimization procedure.} To take maximum advantage of semantic encoding, the feature maps should not be concatenated before the bottleneck layer. We postulate that hooking could be best done at the beginning of the decoder in the target branch, to take maximum advantage of the inherent up-sampling in the decoding path, where the concatenated feature maps can benefit from every skip connection within the target branch. We call this concatenation ``hooking'',  and in order to guarantee pixel-wise alignment in feature maps, we define the spatial resolution of a feature map as $SRF = 2^dr$, where $d$ is the depth in the encoder-decoder model and $r$ is the resolution of the input patch measured in $\mu m/px$.
To define the relative depths were the hooking can take place, we define a SFR ratio between a pair of feature maps as 

\begin{equation}\label{eq:ratio}
\frac{SRF_C}{SRF_T} = 2^{d_C-d_T}\frac{r_C}{r_T}
\end{equation}

where $d_T$ and $d_C$, are the relative depths for the target and context branch, respectively. In practice, hooking can take place when feature maps from both branches comprise the same resolution: $\frac{SRF_C}{SRF_T}=1$.

As a result, the central square region in the feature maps of the context branch at depth $d_C$, are corresponding to the feature maps of the target branch at depth $d_T$. The size of this central square region is equal to the size of feature maps of the target branch because both feature maps comprise the same resolution. To do the actual hooking, simple cropping can be applied, such that context branch feature maps are \emph{pixel aligned} concatenated together with feature maps in the target branch. 

\subsection{Targets and losses}
The goal of HookNet is to predict a segmentation map based on  $P_C$ and $P_T$. Therefore, it can be trained with a single loss backpropagated via the target loss computed for the output of the target branch. Also, the context branch can generate the predictions for the lower resolution patch, and this context error can be used simultaneously with the target loss. {\color{black}For training purpose, we propose a loss function $L = \lambda L_{high} + (1-\lambda) L_{low}$, where $L_{high}$ and $L_{low}$ are pixel-wise categorical cross-entropy for the target and the context branch, respectively, and $\lambda$ controls the importance of each branch.}\


\subsection{Pixel-based-sampling}
Patches are sampled with a particular tissue type, i.e., class label, at the center location of the sampled patch. Due to the sparseness of the ground truth labels, some patches contain less ground truth pixels than other patches. During training, we ensured that every class label is equally represented through the following pixel-based sampling strategy. In the first mini-batch, patches are randomly sampled. In all subsequent mini-batches, patch sampling is guided based on the accumulation of the ground-truth pixels for every class seen in the previous mini-batches. Classes that have a lower amount of pixel accumulation have a higher chance of being sampled to compensate underrepresented classes.


\subsection{Model training setup}\label{training_setup}
Patches were extracted with 284x284x3 in dimensions, and we used a mini-batch size of 12, which allows for two times the number of classes to be in a batch.
Convolutional layers used valid convolutions, $L_2$ regularizer, and the ReLU activation function. Each convolutional layer was followed by batch-normalization. Both branches consisted of a depth of 4 (i.e., 4 down-sampling and 4 up-sampling operations). As mentioned in section \ref{section:branches},  for down- and up-sampling operations, we used 2x2 max-pooling and 2x2 nearest-neighbours followed by a convolutional layer. To predict the soft labels, we used the softmax activation function. {\color{black} The contribution of the losses from the target and context branch can be controlled with a $\lambda$ value. We have tested $\lambda=1$ to ignore the context loss, $\lambda=0.75$ to give more importance to the target branch, $\lambda=0.5$ for equal importance and $\lambda=0.25$ to give more importance to the context loss.} Moreover, we made use of the Adam optimizer with a learning rate of $5x10^{-6}$. We customized the number of filters for all models, such that every model has approximately 50 million parameters. We trained for 200 epochs where each epoch consist of 1000 training steps followed by the calculation of the $F_1$ score on the validation set, which was used to determine the best model. To increase the dataset and account for color changes induced by the variability of staining, we applied spatial, color, noise and stain (\cite{tellez_whole-slide_2018}) augmentations. {\color{black}No stain normalization techniques were used in this work.}

\section{Experiments}

In order to assess HookNet we compared it to five individual U-Nets trained with patches extracted at the following resolutions: 0.5, 1.0, 2.0, 4.0, and 8.0 $\mu m/px$. {\color{black} The models are expressed as U-Net($r_t$) and HookNet($r_t$, $r_c$), where $r_t$ and $r_c$ are the input resolutions for the target and context branch, respectively.}
The aim of HookNet is to output high-resolution segmentation maps, and thereupon the target branch will process input patches extracted at 0.5 $\mu m/px$. For the context branch, we extracted patches at 2.0, or 8.0 $\mu m/px$ {\color{black} (breast only) }, which are the intermediate and extreme resolutions that we tested for the single-resolution models {\color{black} and showed potential value in single resolution performance measures for the breast and lung data (as can be seen in Tables \ref{f1unetsbreast} and \ref{f1unetslung}). We have applied all models to the breast and the lung dataset, with an exception for HookNet trained with resolutions 0.5 and 8.0 $\mu m/px$ on the lung data, because it was evident from the single resolution models (U-Net trained with resolutions 4.0 or 8.0 $\mu m/px$) that no potential information is available at these resolutions for this particular use case.} {\color{black} In the HookNet models, 'Hooking', from the context branch into the target branch takes place at relative depths where the features maps of both branches comprise the same resolution, which is dependent on the input resolutions. Considering the target resolution 0.5 $\mu m/px$, we applied 'Hooking' from depth 2 (the middle) of the context encoder and depth 0 (the end) of the context decoder) into depth 4 (the start or bottleneck) of the target decoder, respectively for the context resolution 2.0 $\mu m/px$ and 8.0 $\mu m/px$} 

To the best of our knowledge, the model proposed by \cite{gu_multi-resolution_2018}, namely MRN, is the most recent model along the same line as HookNet. Therefore, we compared HookNet to MRN. {\color{black}However, HookNet is different from MRN by (1) using 'valid' instead of 'same' convolutions, (2) using an additional branch consisting of an encoder-decoder (which enables multi loss models) instead of a branch with an encoder only and (3)  single upsampling via the decoder of the target branch instead of multiple independent upsamplings.} We instantiated MRN with one extra encoder and used input sizes of 256x256x3. The convolutions in MRN make use of \emph{same} padding, which results in a bigger output size compared to using valid convolutions, therefore allowing for more pixel examples in each output prediction. For this reason and to allow MRN to be trained on a single GPU, we used a mini-batch size of 6 instead of 12. 

{\color{black}
All U-Net models and the HookNet model using a single loss (where lambda$=$1) are trained within approximately 2 days. HookNet trained with the additional contextual loss and MRN, are trained within approximately 2.5 days. We argue that this increase is due to the extra loss in HookNet and the larger size of the feature maps in MRN, which were a result of using 'same' padding. All training times were measured using a GeForce GTX 1080 Ti and 10 CPUs for parallel patch extraction and data augmentation.}


\begin{figure*}[ht]
\centering

\begin{subfigure}{.25\linewidth}
    \includegraphics[width=\linewidth]{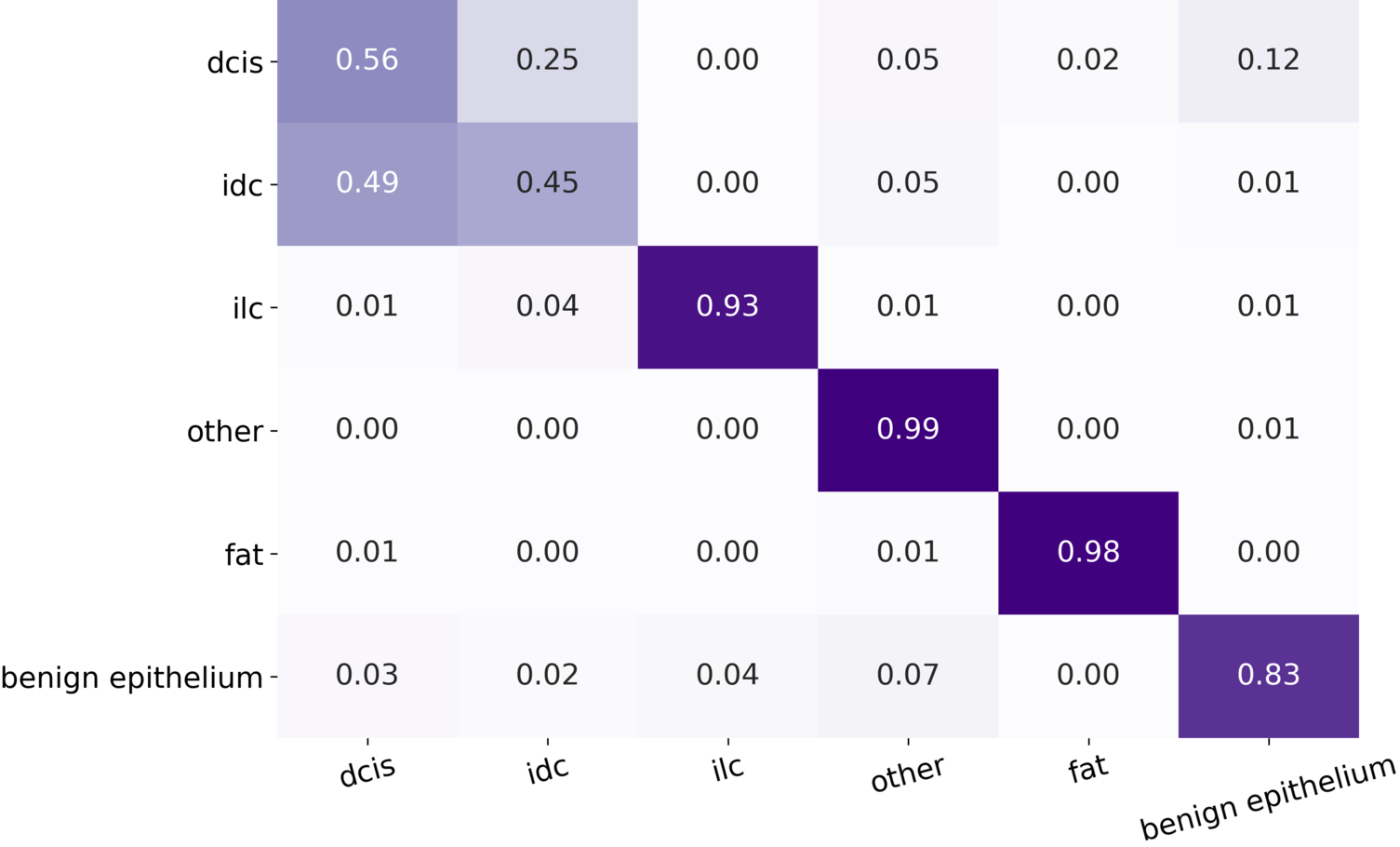}
    \caption{U-Net (0.5 $\mu m/px$)}
\end{subfigure} \qquad
\begin{subfigure}{.25\linewidth}
    \includegraphics[width=\linewidth]{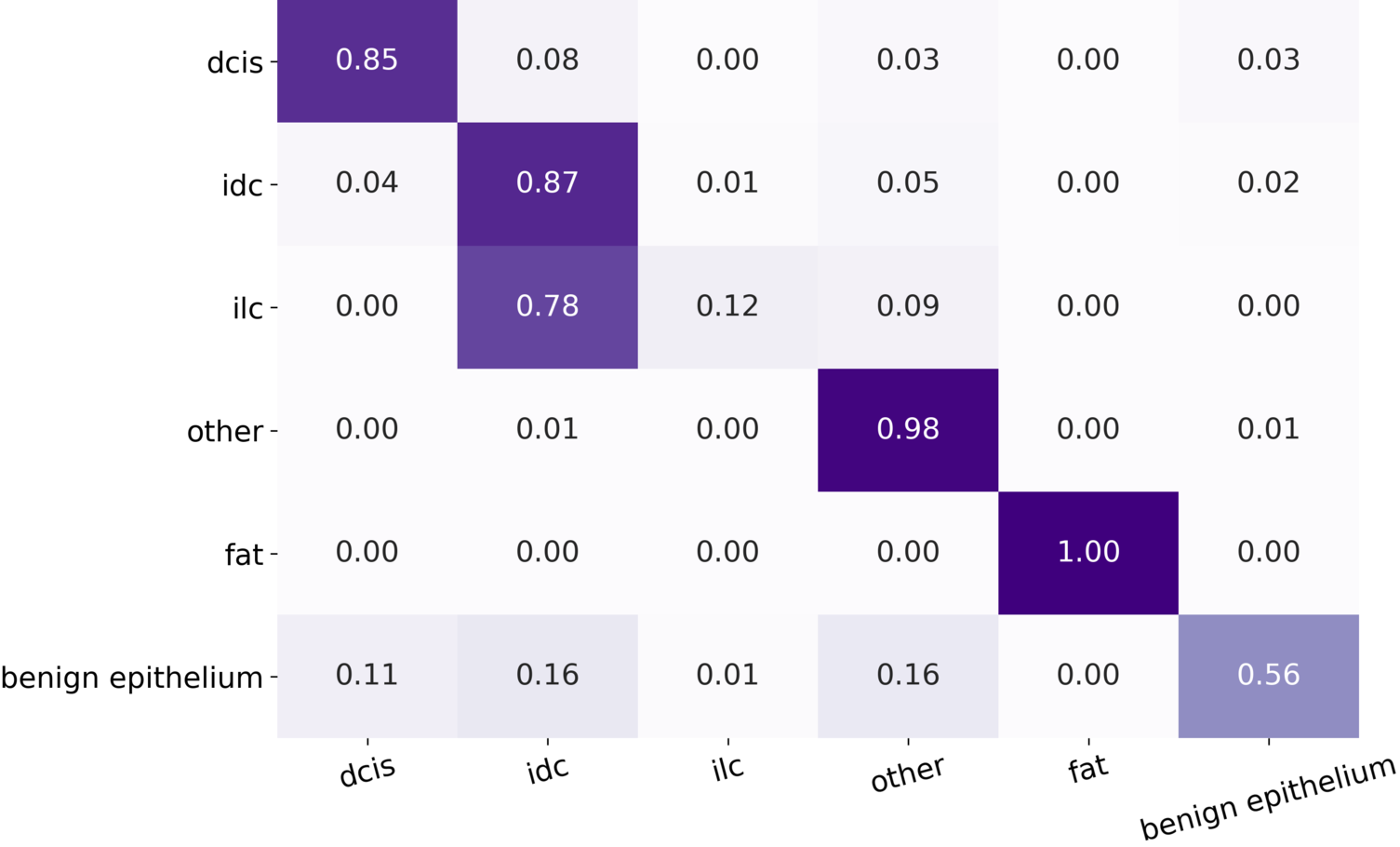}
    \caption{U-Net (8.0 $\mu m/px$)}
\end{subfigure} \qquad
\begin{subfigure}{.25\linewidth}
    \includegraphics[width=\linewidth]{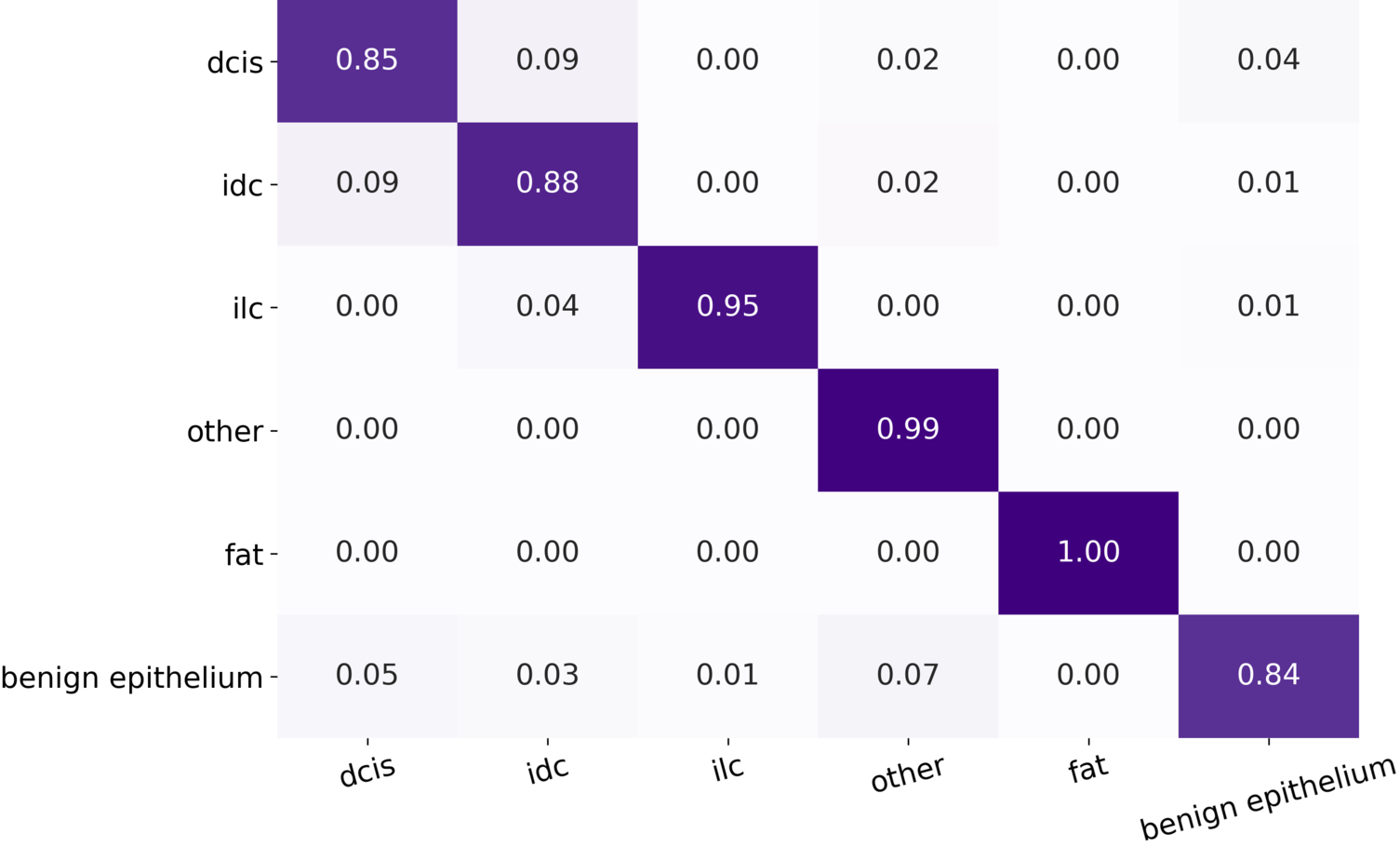}
    \caption{HookNet (0.5, 8.0 $\mu m/px$)}
\end{subfigure} \qquad
\caption{Confusion matrices for models U-Net(0.5), U-Net(8.0) and HookNet(0.5, 8.0,  {\color{black} $\lambda$=0.75}), which were trained on the breast dataset.}
\label{fig:confusion_matrices_breast}
\end{figure*}

\begin{figure*}[ht]
\centering

\begin{subfigure}{.2\linewidth}
    \includegraphics[width=\linewidth]{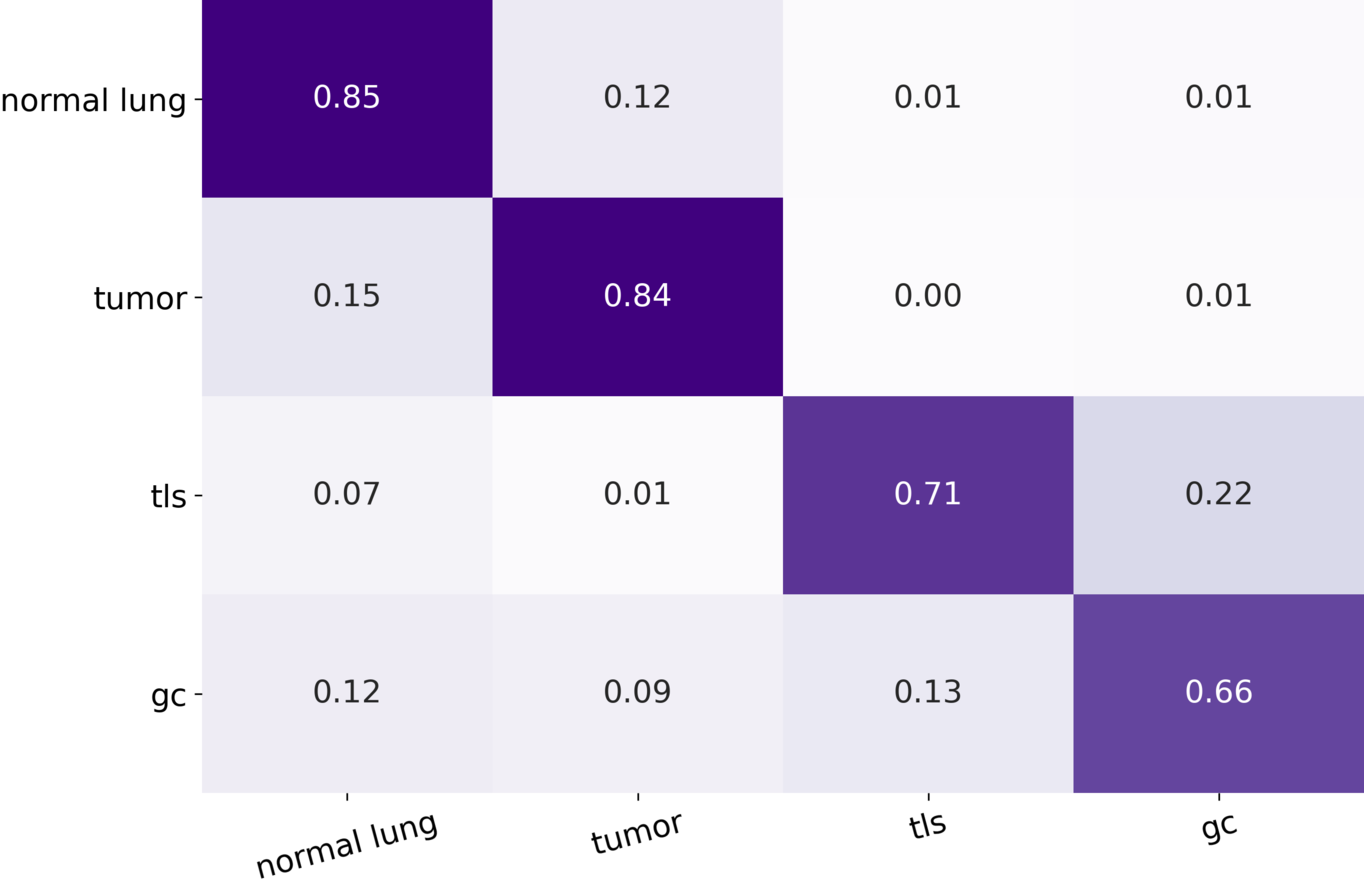}
    \caption{U-Net (0.5 $\mu m/px$)}
\end{subfigure} \qquad
\begin{subfigure}{.2\linewidth}
    \includegraphics[width=\linewidth]{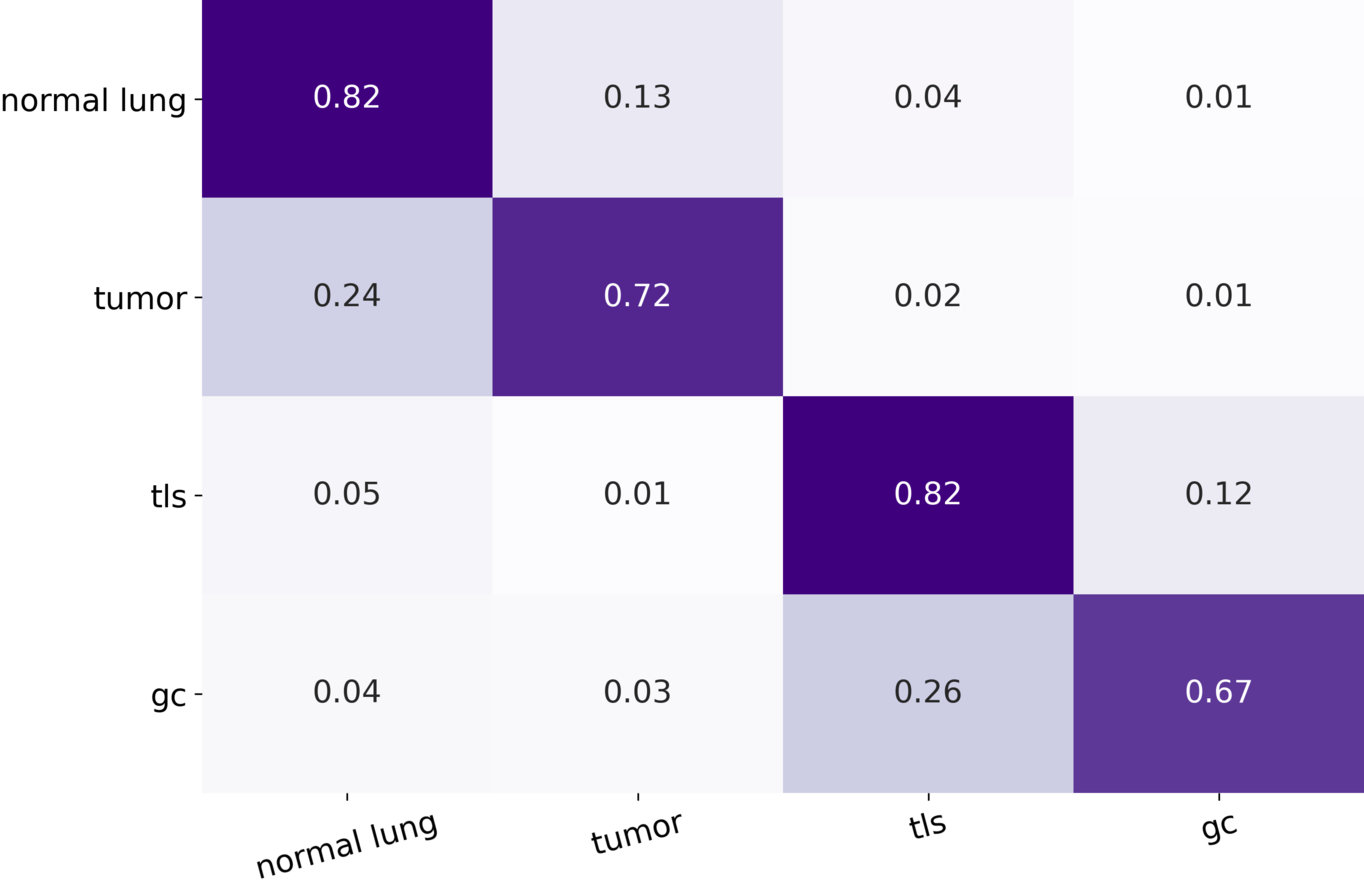}
    \caption{U-Net (2.0 $\mu m/px$)}
\end{subfigure} \qquad
\begin{subfigure}{.2\linewidth}
    \includegraphics[width=\linewidth]{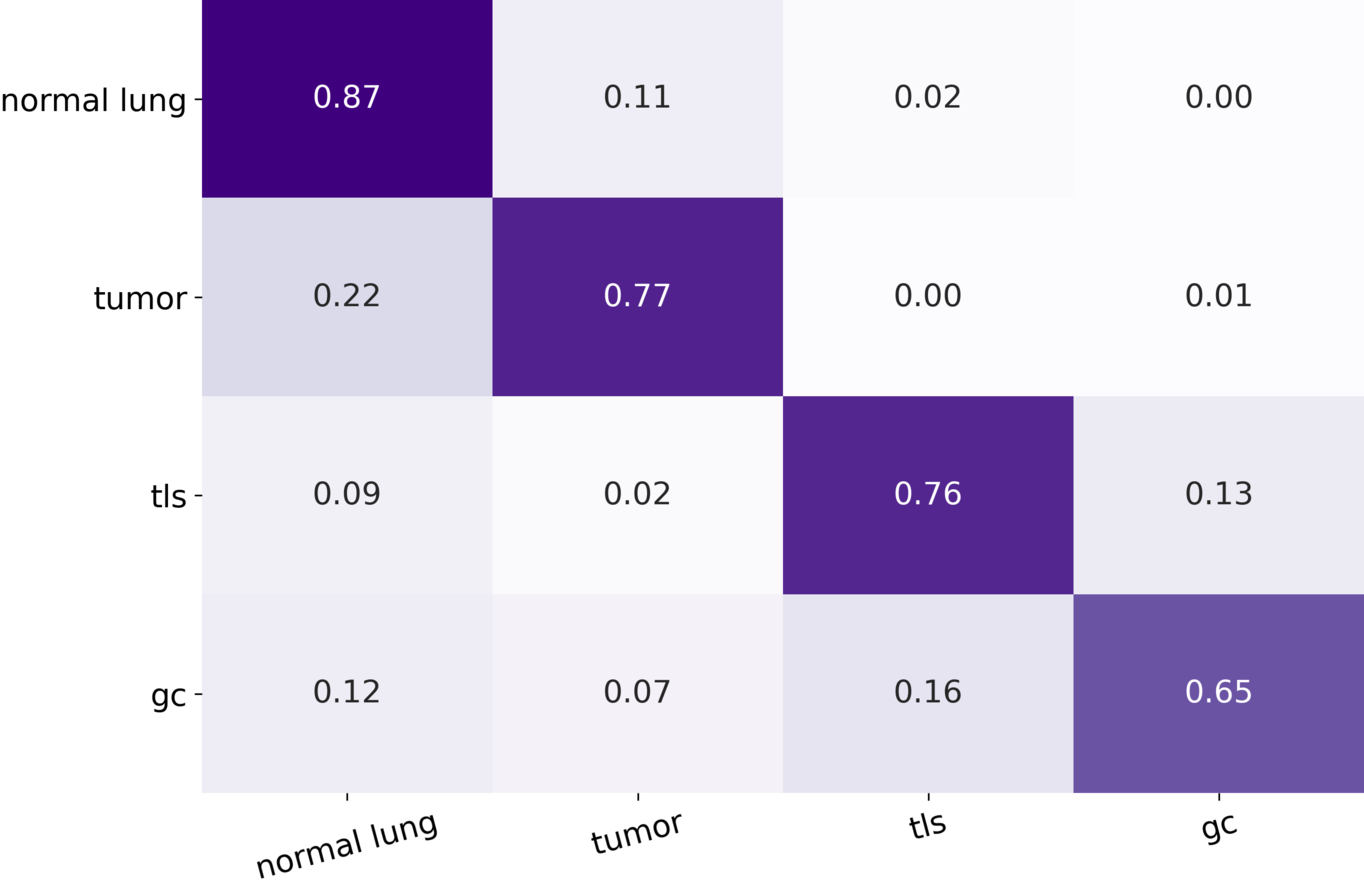}
    \caption{HookNet (0.5, 2.0 $\mu m/px$)}
\end{subfigure} \qquad
\caption{Confusion matrices for models U-Net({\color{black}0.5}), U-Net({\color{black}2.0}) and HookNet({\color{black}0.5, 2.0}, {\color{black} $\lambda$=1}), which were trained on the lung dataset.}
\label{fig:confusion_matrices_lung}
\end{figure*}

\section{Results}\label{results}


Quantitative performance, for the breast data set, in terms of $F_1$ score for each considered class as well as an overall Macro $F_1$ \citep{haghighi_pycm:_2018} are reported in Table {\color{black} \ref{f1unetsbreast}, for all U-Net models and for each considered resolution. Quantitative performance for all models with target resolution $0.5 \mu m/px$ (i.e., U-Net(0.5), MRN and HookNet) are reported in Table \ref{f1breast}. Likewise, for the lung dataset, quantitative performance are reported in Table \ref{f1unetslung} for all U-Net models for each considered resolution and in Table \ref{f1lung} quantitative performance are reported for all models with target resolution $0.5 \mu m/px$ (i.e., U-Net(0.5), MRN and HookNet).}

\begin{figure}[t]
\captionsetup[subfigure]{labelformat=empty, position=above}
\centering

\begin{subfigure}[b]{.14\linewidth}
    \caption{\centering Tissue}
    \includegraphics[width=\linewidth]{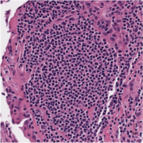}
\end{subfigure}
\begin{subfigure}[b]{.14\linewidth}
    \caption{\centering Ground truth}
    \includegraphics[width=\linewidth]{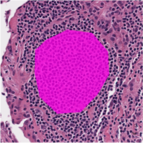}
\end{subfigure}
\begin{subfigure}[b]{.14\linewidth}
    \caption{\centering U-Net\hspace{\textwidth}\tiny0.5}
\includegraphics[width=\linewidth]{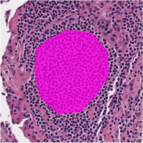}
\end{subfigure}
\begin{subfigure}[b]{.14\linewidth}
    \caption{\centering U-Net\hspace{\textwidth}\tiny2.0}
\includegraphics[width=\linewidth]{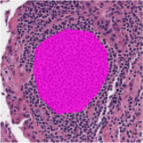}
\end{subfigure}
\begin{subfigure}[b]{.14\linewidth}
    \caption{\centering MRN\hspace{\textwidth}\tiny0.5, 2.0}
\includegraphics[width=\linewidth]{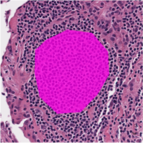}
\end{subfigure}
\begin{subfigure}[b]{.14\linewidth}
    \caption{\centering HookNet\hspace{\textwidth}\tiny0.5, 2.0}
\includegraphics[width=\linewidth]{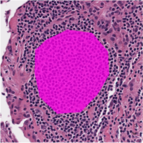}
\end{subfigure}

\vspace{1mm}
\begin{subfigure}[b]{.14\linewidth}
    \includegraphics[width=\linewidth]{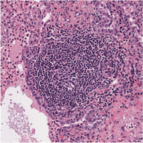}
\end{subfigure}
\begin{subfigure}[b]{.14\linewidth}
    \includegraphics[width=\linewidth]{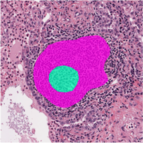}
\end{subfigure}
\begin{subfigure}[b]{.14\linewidth}
    \includegraphics[width=\linewidth]{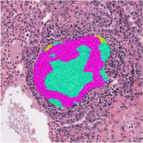}
\end{subfigure}
\begin{subfigure}[b]{.14\linewidth}
    \includegraphics[width=\linewidth]{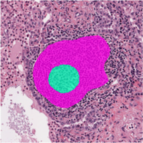}
\end{subfigure}
\begin{subfigure}[b]{.14\linewidth}
    \includegraphics[width=\linewidth]{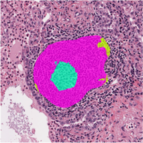}
\end{subfigure}
\begin{subfigure}[b]{.14\linewidth}
    \includegraphics[width=\linewidth]{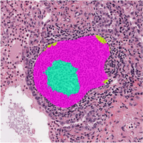}
\end{subfigure}

\vspace{1mm}

\begin{subfigure}[b]{.14\linewidth}
    \includegraphics[width=\linewidth]{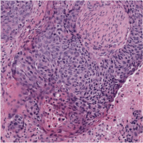}
\end{subfigure}
\begin{subfigure}[b]{.14\linewidth}
    \includegraphics[width=\linewidth]{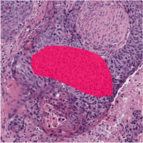}
\end{subfigure}
\begin{subfigure}[b]{.14\linewidth}
    \includegraphics[width=\linewidth]{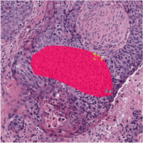}
\end{subfigure}
\begin{subfigure}[b]{.14\linewidth}
    \includegraphics[width=\linewidth]{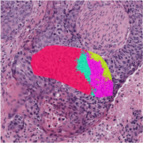}
\end{subfigure}
\begin{subfigure}[b]{.14\linewidth}
    \includegraphics[width=\linewidth]{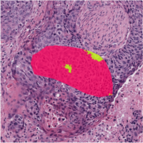}
\end{subfigure}
\begin{subfigure}[b]{.14\linewidth}
    \includegraphics[width=\linewidth]{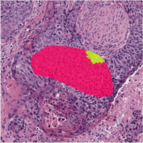}
\end{subfigure}

\vspace{1mm}

\begin{subfigure}[b]{.14\linewidth}
    \includegraphics[width=\linewidth]{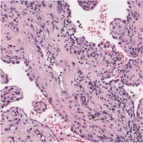}
\end{subfigure}
\begin{subfigure}[b]{.14\linewidth}
    \includegraphics[width=\linewidth]{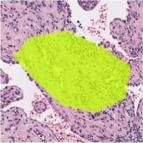}
\end{subfigure}
\begin{subfigure}[b]{.14\linewidth}
    \includegraphics[width=\linewidth]{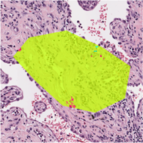}
\end{subfigure}
\begin{subfigure}[b]{.14\linewidth}
    \includegraphics[width=\linewidth]{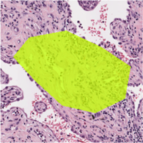}
\end{subfigure}
\begin{subfigure}[b]{.14\linewidth}
    \includegraphics[width=\linewidth]{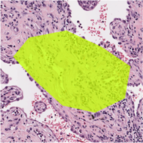}
\end{subfigure}
\begin{subfigure}[b]{.14\linewidth}
    \includegraphics[width=\linewidth]{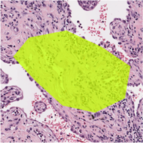}
\end{subfigure}

\vspace{6mm}

\begin{subfigure}[b]{.14\linewidth}
    \includegraphics[width=\linewidth]{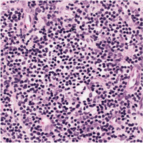}
\end{subfigure}
\begin{subfigure}[b]{.14\linewidth}
    \includegraphics[width=\linewidth]{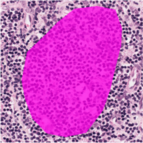}
\end{subfigure}
\begin{subfigure}[b]{.14\linewidth}
    \includegraphics[width=\linewidth]{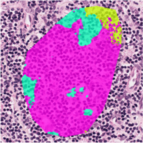}
\end{subfigure}
\begin{subfigure}[b]{.14\linewidth}
    \includegraphics[width=\linewidth]{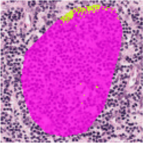}
\end{subfigure}
\begin{subfigure}[b]{.14\linewidth}
    \includegraphics[width=\linewidth]{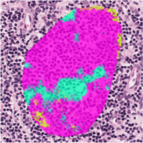}
\end{subfigure}
\begin{subfigure}[b]{.14\linewidth}
    \includegraphics[width=\linewidth]{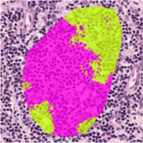}
\end{subfigure}

\vspace{1mm}

\begin{subfigure}[b]{.14\linewidth}
    \includegraphics[width=\linewidth]{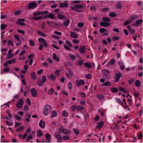}
\end{subfigure}
\begin{subfigure}[b]{.14\linewidth}
    \includegraphics[width=\linewidth]{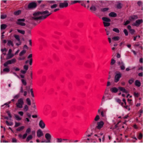}
\end{subfigure}
\begin{subfigure}[b]{.14\linewidth}
    \includegraphics[width=\linewidth]{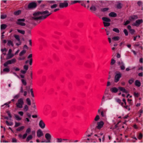}
\end{subfigure}
\begin{subfigure}[b]{.14\linewidth}
    \includegraphics[width=\linewidth]{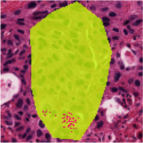}
\end{subfigure}
\begin{subfigure}[b]{.14\linewidth}
    \includegraphics[width=\linewidth]{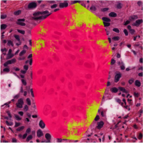}
\end{subfigure}
\begin{subfigure}[b]{.14\linewidth}
    \includegraphics[width=\linewidth]{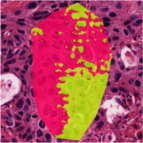}
\end{subfigure}

\begin{subfigure}[t]{.05\linewidth}
\includegraphics[width=\linewidth]{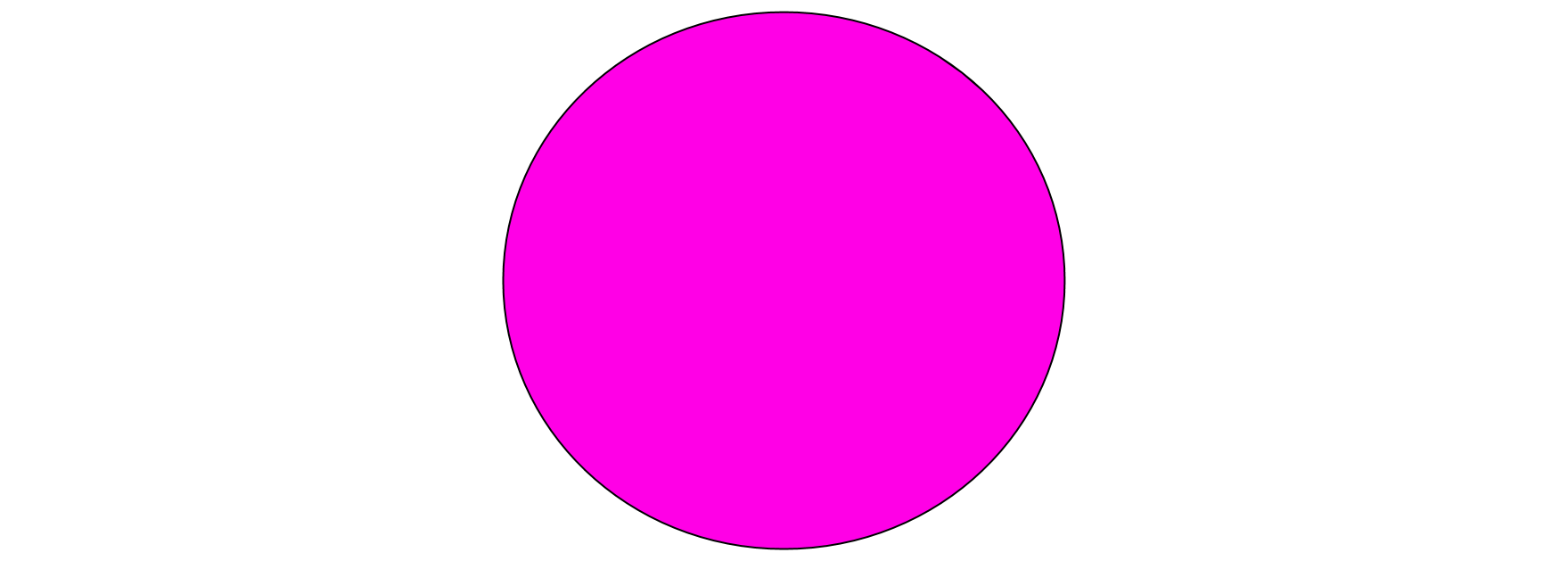}
\end{subfigure}
\hspace{-2mm}
\begin{subfigure}[t]{.10\linewidth}
\tiny TLS
\end{subfigure}
\begin{subfigure}[t]{.05\linewidth}
\includegraphics[width=\linewidth]{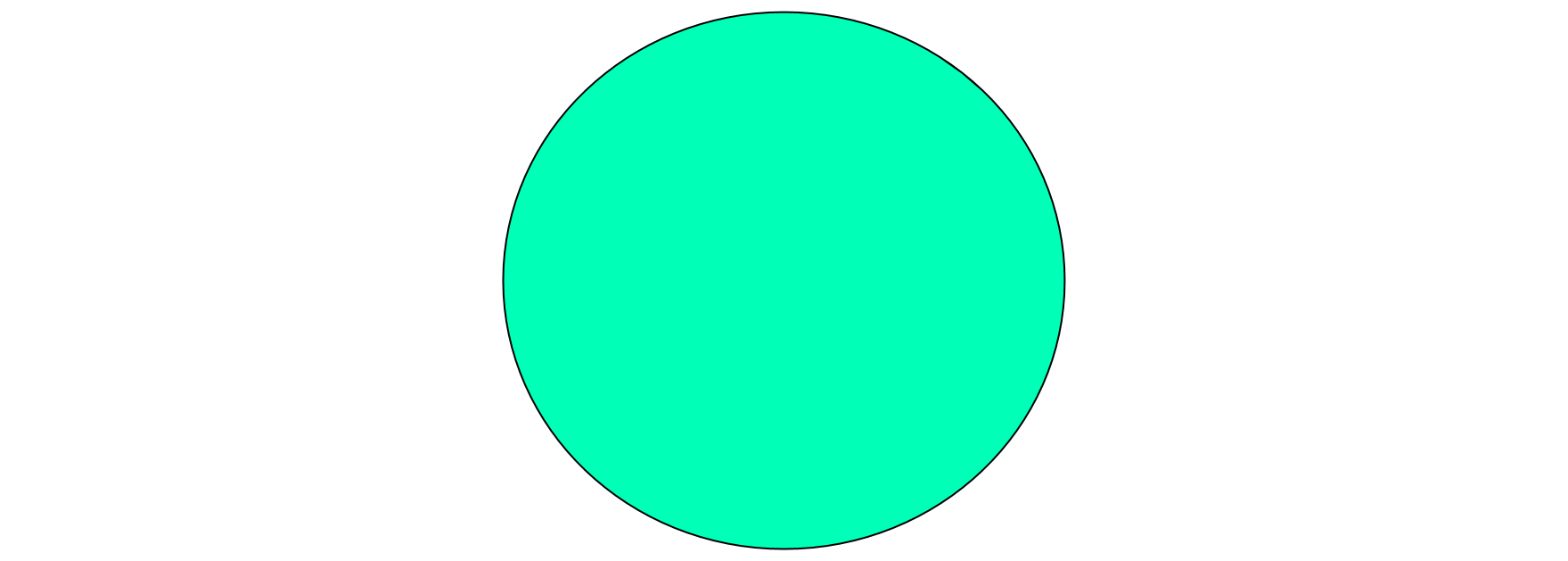}
\end{subfigure}
\hspace{-2mm}
\begin{subfigure}[t]{.10\linewidth}
\tiny GC
\end{subfigure}
\begin{subfigure}[t]{.05\linewidth}
\includegraphics[width=\linewidth]{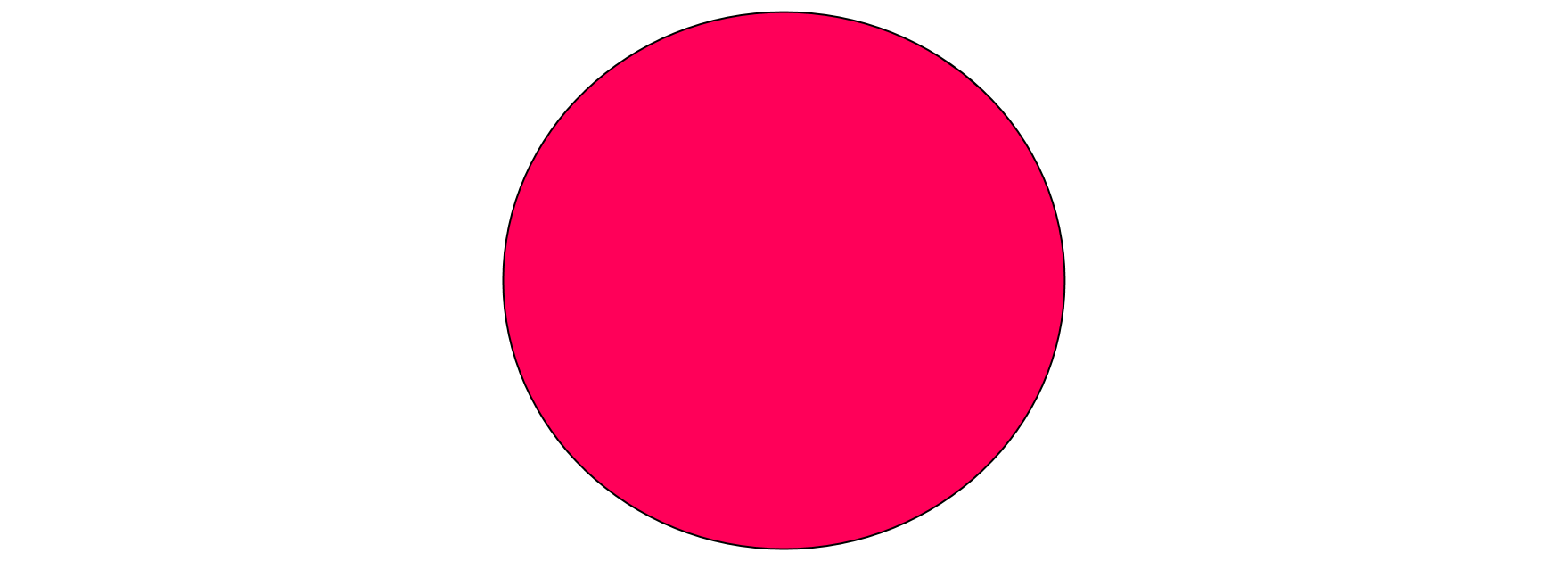}
\end{subfigure}
\hspace{-2mm}
\begin{subfigure}[t]{.10\linewidth}
\tiny Tumor
\end{subfigure}
\begin{subfigure}[t]{.05\linewidth}
\includegraphics[width=\linewidth]{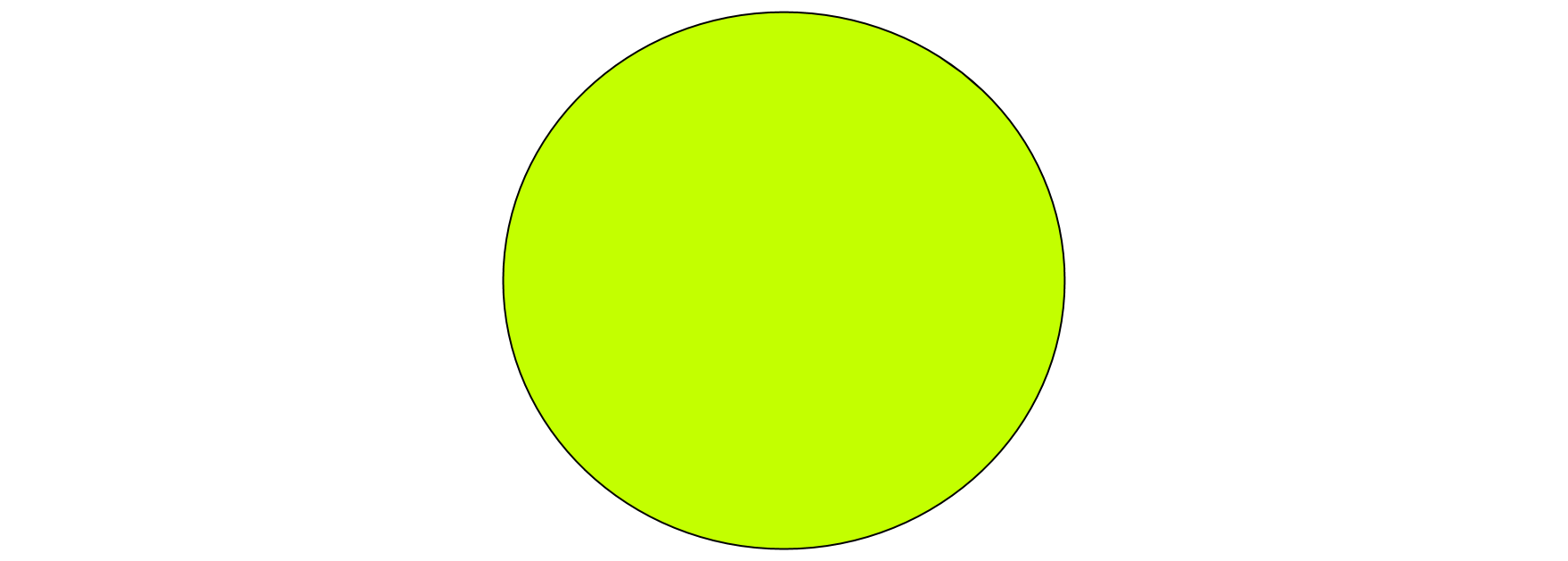}
\end{subfigure}
\hspace{-2mm}
\begin{subfigure}[t]{.10\linewidth}
\tiny Other
\end{subfigure}

\caption{Segmentation results on lung tissue for TLS and GC, Tumor, and Other. {\color{black} HookNet results are shown for $\lambda$=1.0. The last two rows focus on failure examples for HookNet}}
\label{fig:lung_performance}
\end{figure}

Confusion matrices for U-Net and HookNet models for breast and lung test sets are depicted in Figure \ref{fig:confusion_matrices_breast} and Figure \ref{fig:confusion_matrices_lung}, respectively.
Finally, visual results are shown for each class of breast and lung tissue in Figure \ref{fig:breast_performance} and Figure \ref{fig:lung_performance} respectively.

\subsection{Single-resolution models}
Experimental results of single-resolution U-Net on DCIS and ILC confirm our initial hypothesis, namely an increase in performance that correlates with increase of context (microns per pixel) for DCIS (e.g., from $F_1$=0.47 at 0.5 $\mu m/px$ to $F_1$=0.86 at 8.0 $\mu m/px$), and a completely opposite trend for ILC (e.g., from $F_1$=0.85 at 0.5 $\mu m/px$ to $F_1$=0.20 at 8.0 $\mu m/px$), corroborating the needs for a multi-resolution model such as HookNet.
As expected, the lack of context causes confusion between DCIS and IDC in U-Net(0.5), where breast duct structures are not visible due to the limited field of view, whereas the lack of details causes confusion between ILC and IDC in U-Net(8.0) (see Figure \ref{fig:confusion_matrices_breast}), where all loose tumor cells are interpreted as part of a single bulk.

The highest performance in IDC and benign breast epithelium where observed at relatively intermediate resolutions.
The performance of segmentation of fatty tissue is comparable in every model, and the performance of segmenting other tissue decreases when using relatively low resolutions (4.0 and 8.0 $\mu m/px$) or high-resolution (0.5 $\mu m/px$).

For lung tissue, we observed an increase in performance that correlates with an increase in context and decrease in resolution. This is mostly due to an increase in $F_1$ score for {\color{black}GC in U-Net(2.0)}, similar to the increase observed for DCIS, whereas lack of details causes confusion between {\color{black}Tumor and Other in U-Net(2.0)}, similar to what observed for ILC in breast tissue.

\subsection{Multi-resolution models}
\begin{figure}[t]
\captionsetup[subfigure]{labelformat=empty, position=above}
\centering
\begin{subfigure}[b]{.14\linewidth}
    \caption{\centering Tissue}
    \includegraphics[width=\linewidth]{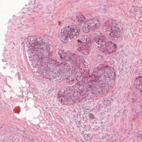}
\end{subfigure}
\begin{subfigure}[b]{.14\linewidth}
    \caption{\centering Ground truth}
    \includegraphics[width=\linewidth]{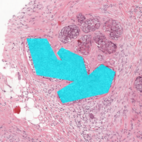}
\end{subfigure}
\begin{subfigure}[b]{.14\linewidth}
    \caption{\centering U-Net\hspace{\textwidth}\tiny0.5}
    \includegraphics[width=\linewidth]{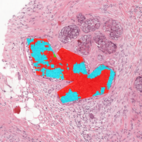}
\end{subfigure}
\begin{subfigure}[b]{.14\linewidth}
    \caption{\centering U-Net\hspace{\textwidth}\tiny8.0}
    \includegraphics[width=\linewidth]{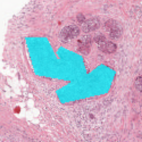}
\end{subfigure}
\begin{subfigure}[b]{.14\linewidth}
    \caption{\centering MRN\hspace{\textwidth}\tiny0.5, 8.0}
    \includegraphics[width=\linewidth]{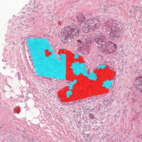}
\end{subfigure}
\begin{subfigure}[b]{.14\linewidth}
    \caption{\centering HookNet\hspace{\textwidth}\tiny0.5, 8.0}
    \includegraphics[width=\linewidth]{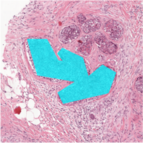}
\end{subfigure}

\vspace{1mm}

\begin{subfigure}[b]{.14\linewidth}
    \includegraphics[width=\linewidth]{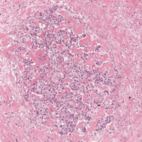}
\end{subfigure}
\begin{subfigure}[b]{.14\linewidth}
    \includegraphics[width=\linewidth]{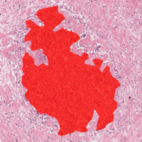}
\end{subfigure}
\begin{subfigure}[b]{.14\linewidth}
\includegraphics[width=\linewidth]{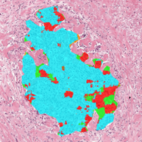}
\end{subfigure}
\begin{subfigure}[b]{.14\linewidth}
    \includegraphics[width=\linewidth]{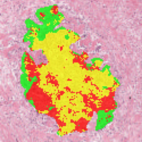}
\end{subfigure}
\begin{subfigure}[b]{.14\linewidth}
    \includegraphics[width=\linewidth]{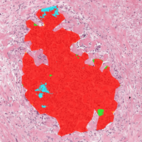}
\end{subfigure}
\begin{subfigure}[b]{.14\linewidth}
    \includegraphics[width=\linewidth]{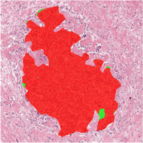}
\end{subfigure}

\vspace{1mm}

\begin{subfigure}[b]{.14\linewidth}
    \includegraphics[width=\linewidth]{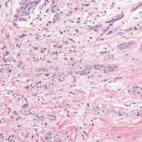}
\end{subfigure}
\begin{subfigure}[b]{.14\linewidth}
    \includegraphics[width=\linewidth]{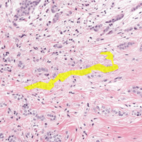}
\end{subfigure}
\begin{subfigure}[b]{.14\linewidth}
    \includegraphics[width=\linewidth]{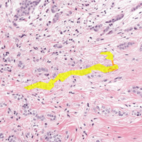}
\end{subfigure}
\begin{subfigure}[b]{.14\linewidth}
    \includegraphics[width=\linewidth]{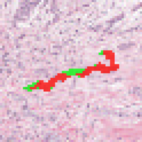}
\end{subfigure}
\begin{subfigure}[b]{.14\linewidth}
    \includegraphics[width=\linewidth]{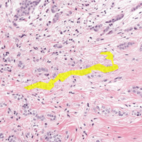}
\end{subfigure}
\begin{subfigure}[b]{.14\linewidth}
    \includegraphics[width=\linewidth]{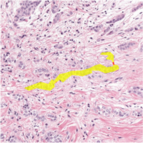}
\end{subfigure}

\vspace{1mm}

\begin{subfigure}[b]{.14\linewidth}
\includegraphics[width=\linewidth]{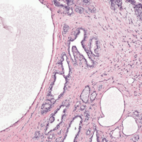}
\end{subfigure}
\begin{subfigure}[b]{.14\linewidth}
\includegraphics[width=\linewidth]{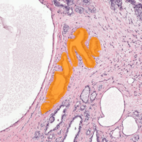}
\end{subfigure}
\begin{subfigure}[b]{.14\linewidth}
\includegraphics[width=\linewidth]{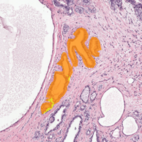}
\end{subfigure}
\begin{subfigure}[b]{.14\linewidth}
\includegraphics[width=\linewidth]{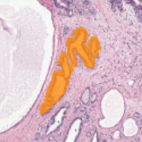}
\end{subfigure}
\begin{subfigure}[b]{.14\linewidth}
\includegraphics[width=\linewidth]{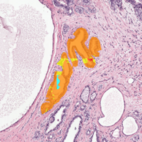}
\end{subfigure}
\begin{subfigure}[b]{.14\linewidth}
\includegraphics[width=\linewidth]{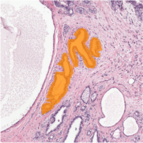}
\end{subfigure}

\vspace{1mm}

\begin{subfigure}[b]{.14\linewidth}
    \includegraphics[width=\linewidth]{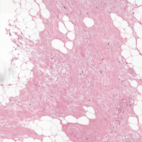}
\end{subfigure}
\begin{subfigure}[b]{.14\linewidth}
    \includegraphics[width=\linewidth]{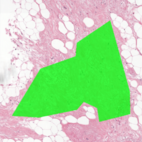}
\end{subfigure}
\begin{subfigure}[b]{.14\linewidth}
    \includegraphics[width=\linewidth]{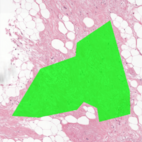}
\end{subfigure}
\begin{subfigure}[b]{.14\linewidth}
    \includegraphics[width=\linewidth]{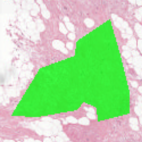}
\end{subfigure}
\begin{subfigure}[b]{.14\linewidth}
    \includegraphics[width=\linewidth]{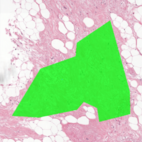}
\end{subfigure}
\begin{subfigure}[b]{.14\linewidth}
    \includegraphics[width=\linewidth]{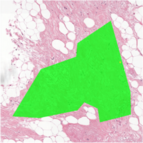}
\end{subfigure}

\vspace{1mm}

\begin{subfigure}[b]{.14\linewidth}
\includegraphics[width=\linewidth]{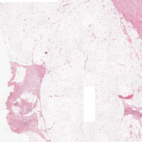}
\end{subfigure}
\begin{subfigure}[b]{.14\linewidth}
\includegraphics[width=\linewidth]{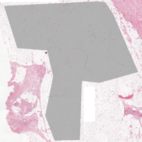}
\end{subfigure}
\begin{subfigure}[b]{.14\linewidth}
\includegraphics[width=\linewidth]{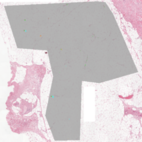}
\end{subfigure}
\begin{subfigure}[b]{.14\linewidth}
\includegraphics[width=\linewidth]{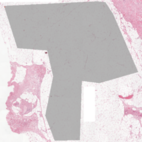}
\end{subfigure}
\begin{subfigure}[b]{.14\linewidth}
\includegraphics[width=\linewidth]{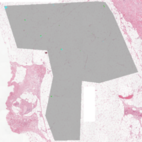}
\end{subfigure}
\begin{subfigure}[b]{.14\linewidth}
\includegraphics[width=\linewidth]{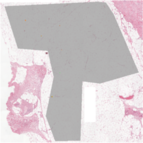}
\end{subfigure}

\vspace{6mm}

\begin{subfigure}[b]{.14\linewidth}
\includegraphics[width=\linewidth]{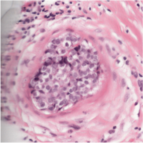}
\end{subfigure}
\begin{subfigure}[b]{.14\linewidth}
\includegraphics[width=\linewidth]{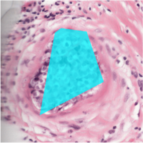}
\end{subfigure}
\begin{subfigure}[b]{.14\linewidth}
\includegraphics[width=\linewidth]{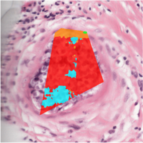}
\end{subfigure}
\begin{subfigure}[b]{.14\linewidth}
\includegraphics[width=\linewidth]{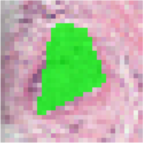}
\end{subfigure}
\begin{subfigure}[b]{.14\linewidth}
\includegraphics[width=\linewidth]{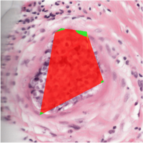}
\end{subfigure}
\begin{subfigure}[b]{.14\linewidth}
\includegraphics[width=\linewidth]{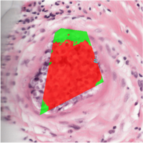}
\end{subfigure}

\vspace{1mm}

\begin{subfigure}[b]{.14\linewidth}
\includegraphics[width=\linewidth]{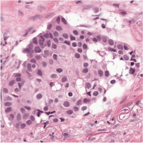}
\end{subfigure}
\begin{subfigure}[b]{.14\linewidth}
\includegraphics[width=\linewidth]{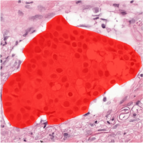}
\end{subfigure}
\begin{subfigure}[b]{.14\linewidth}
\includegraphics[width=\linewidth]{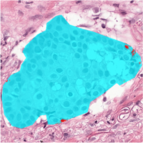}
\end{subfigure}
\begin{subfigure}[b]{.14\linewidth}
\includegraphics[width=\linewidth]{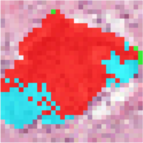}
\end{subfigure}
\begin{subfigure}[b]{.14\linewidth}
\includegraphics[width=\linewidth]{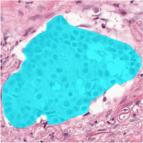}
\end{subfigure}
\begin{subfigure}[b]{.14\linewidth}
\includegraphics[width=\linewidth]{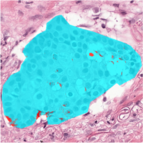}
\end{subfigure}

\vspace{1mm}

\begin{subfigure}[b]{.14\linewidth}
\includegraphics[width=\linewidth]{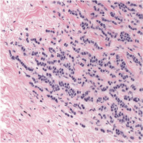}
\end{subfigure}
\begin{subfigure}[b]{.14\linewidth}
\includegraphics[width=\linewidth]{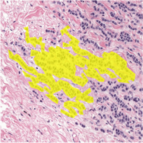}
\end{subfigure}
\begin{subfigure}[b]{.14\linewidth}
\includegraphics[width=\linewidth]{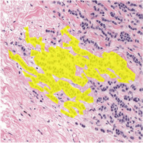}
\end{subfigure}
\begin{subfigure}[b]{.14\linewidth}
\includegraphics[width=\linewidth]{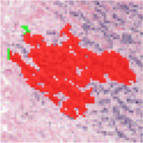}
\end{subfigure}
\begin{subfigure}[b]{.14\linewidth}
\includegraphics[width=\linewidth]{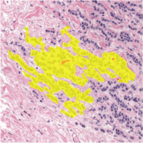}
\end{subfigure}
\begin{subfigure}[b]{.14\linewidth}
\includegraphics[width=\linewidth]{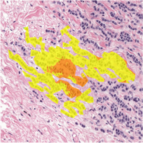}
\end{subfigure}

\begin{subfigure}[t]{.05\linewidth}
\includegraphics[width=\linewidth]{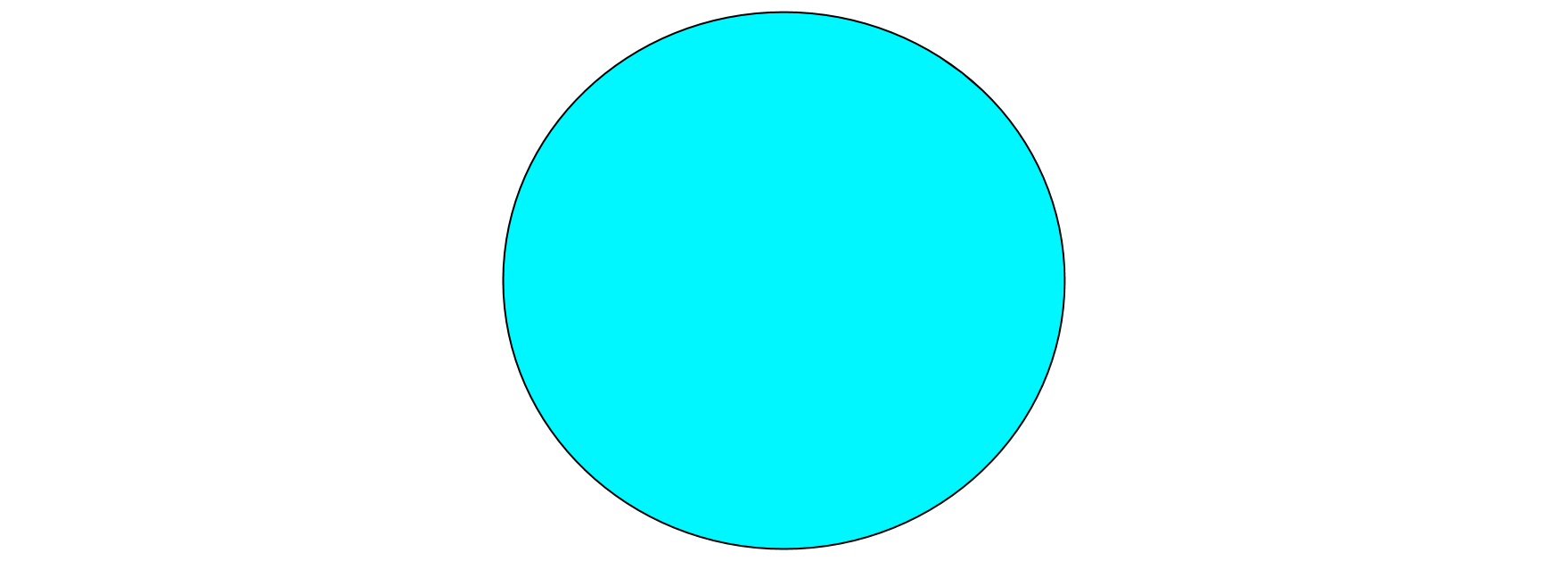}
\end{subfigure}
\hspace{-2mm}
\begin{subfigure}[t]{.10\linewidth}
\tiny DCIS
\end{subfigure}
\begin{subfigure}[t]{.05\linewidth}
\includegraphics[width=\linewidth]{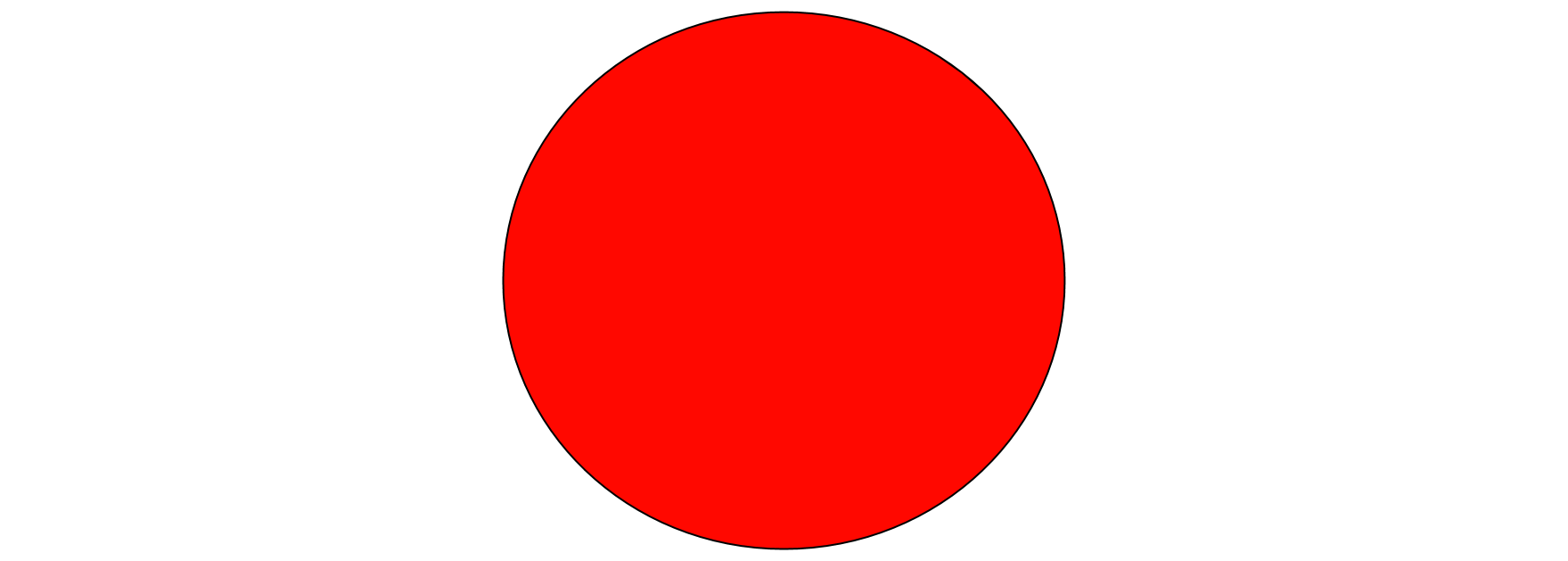}
\end{subfigure}
\hspace{-2mm}
\begin{subfigure}[t]{.10\linewidth}
\tiny IDC
\end{subfigure}
\begin{subfigure}[t]{.05\linewidth}
\includegraphics[width=\linewidth]{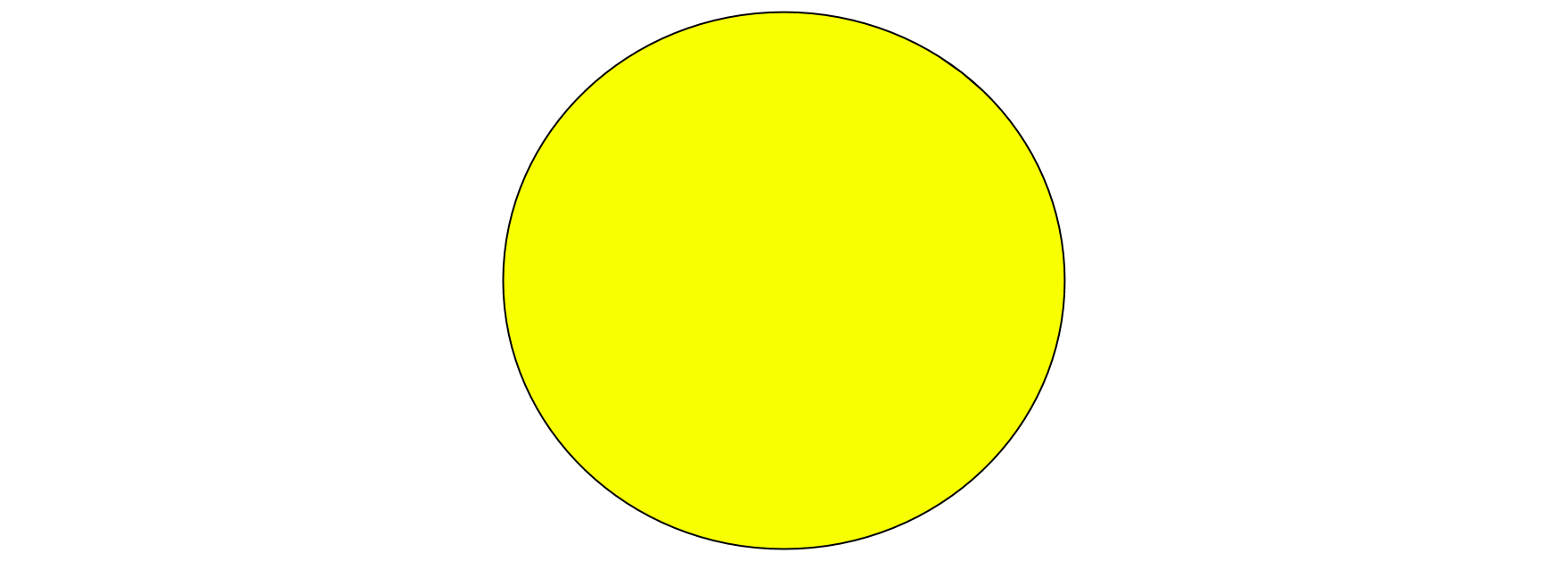}
\end{subfigure}
\hspace{-2mm}
\begin{subfigure}[t]{.10\linewidth}
\tiny ILC
\end{subfigure}
\begin{subfigure}[t]{.05\linewidth}
\includegraphics[width=\linewidth]{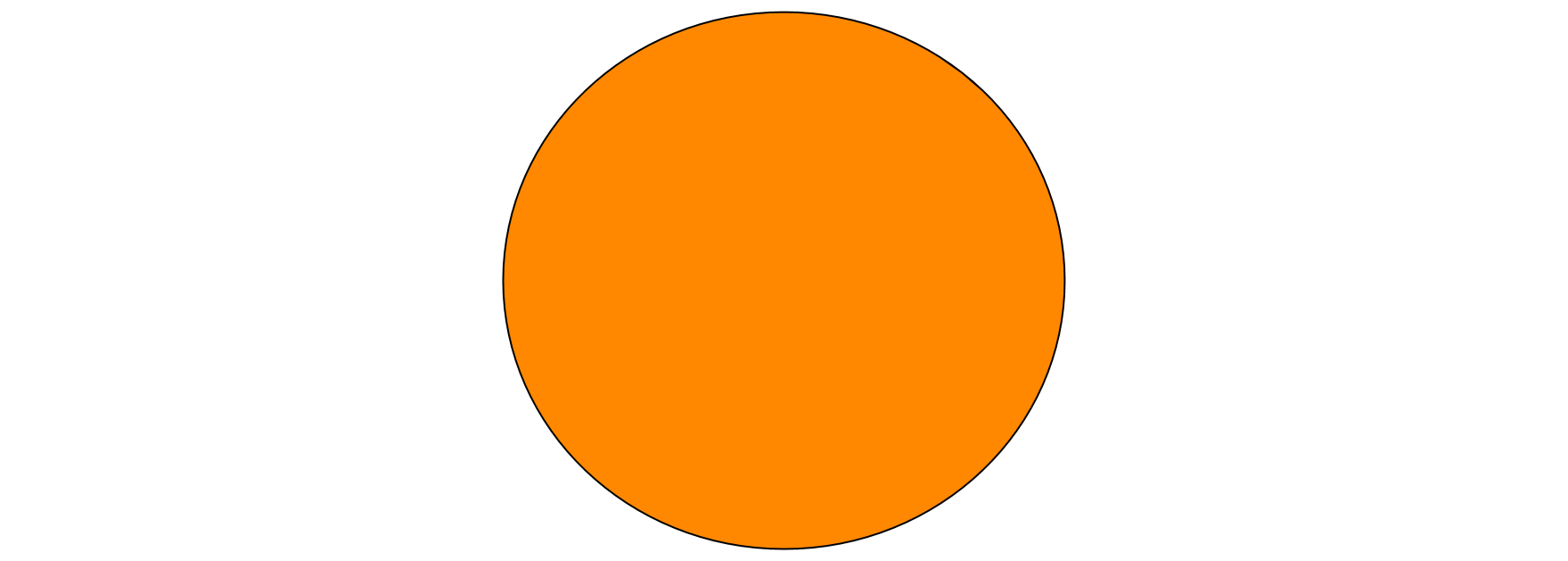}
\end{subfigure}
\hspace{-2mm}
\begin{subfigure}[t]{.10\linewidth}
\tiny Benign\\Epithelium
\end{subfigure}
\begin{subfigure}[t]{.05\linewidth}
\includegraphics[width=\linewidth]{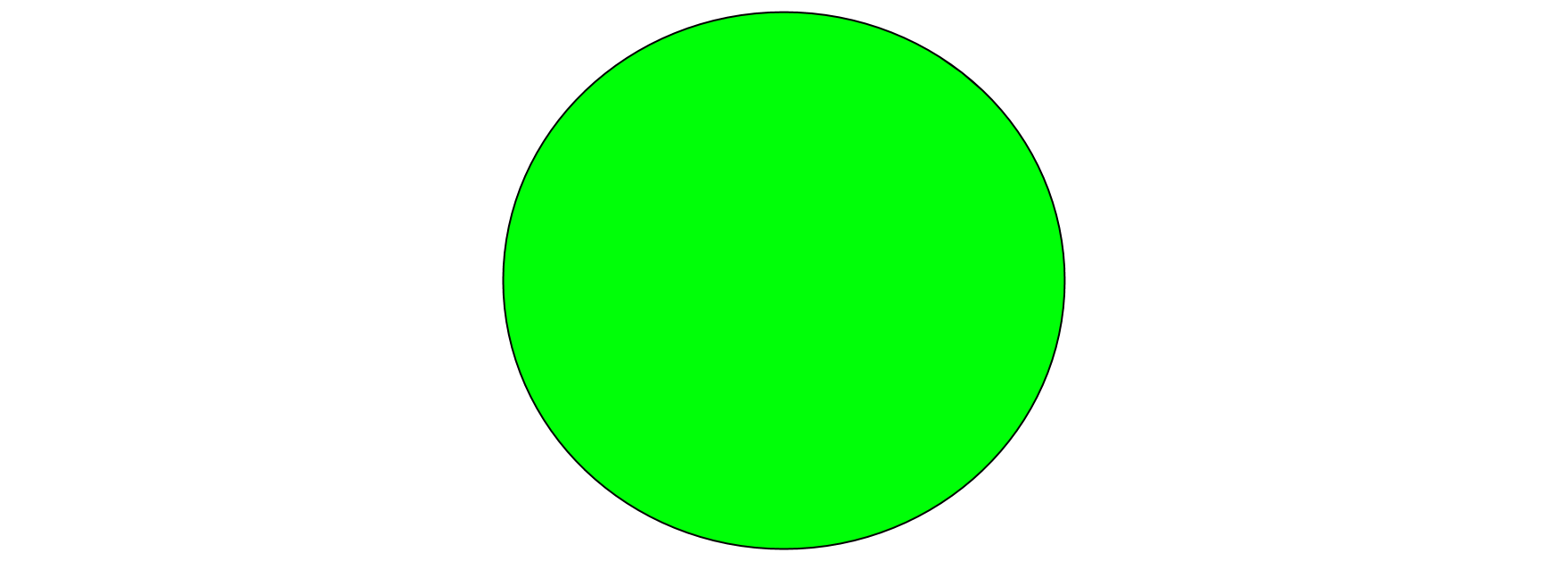}
\end{subfigure}
\hspace{-2mm}
\begin{subfigure}[t]{.10\linewidth}
\tiny Other
\end{subfigure}
\begin{subfigure}[t]{.05\linewidth}
\includegraphics[width=\linewidth]{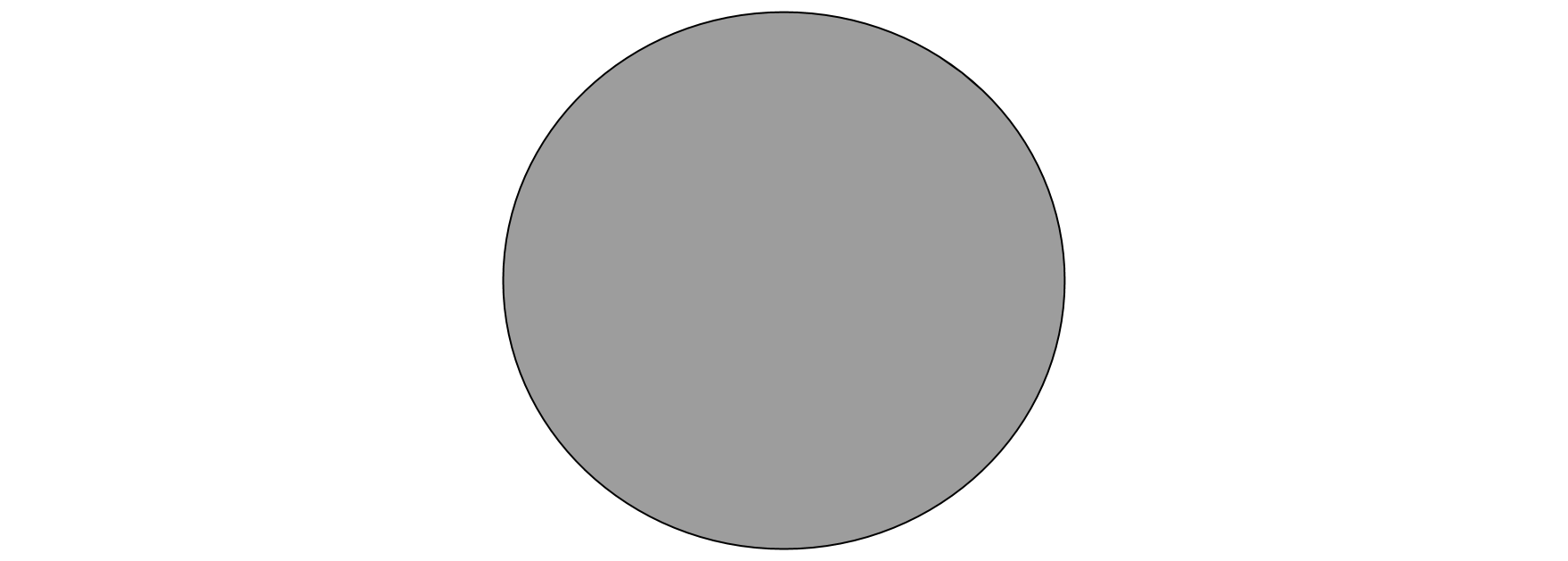}
\end{subfigure}
\hspace{-2mm}
\begin{subfigure}[t]{.10\linewidth}
\tiny Fat
\end{subfigure}

\caption{Segmentation results on breast tissue shown for DCIS, IDC, ILC, Benign epithelium, Other and Fat. {\color{black}HookNet results are shown for $\lambda$=0.75. The last three rows focus on failure examples of HookNet}.}
\label{fig:breast_performance}
\end{figure}

In breast tissue segmentation, performance of HookNet strongly depends on which fields of view are combined. We obtained the best results with an overall $F_1$ score of  {\color{black}0.91} for HookNet(0.5, 8.0) {\color{black}with $\lambda$=0.75}, which substantially differs from HookNet(0.5, 2.0). HookNet(0.5, 8.0) shows an overall increase in all tissue types for the output resolution 0.5 $\mu m/px$ except for a small decrease in performance for DCIS of U-Net trained with patches at resolution 8.0 $\mu m/px$, and improves the performance on IDC, mostly due to an improvement in ILC segmentation, which likely increases the precision of the model for IDC.
Note that DCIS is the only class where U-Net working at the lowest considered resolution gives the best performance ($F_1$=0.86). However, that same U-Net has a dismal $F_1$ score of 0.2 for ILC.
HookNet(0.5, 8.0) processes the same low-resolution input, but increases $F_1$ score for ILC by 0.66 compared to U-Net(8.0), and at the same time increases $F_1$ score for DCIS by 0.37 compared to U-Net(0.5).
In general, HookNet(0.5, 8.0) improves $F_1$ score for all classes compared to U-Net(0.5) and to U-Net(8.0), except for a small difference of 0.02 $F_1$ score in DCIS segmentation.
As for single U-Net models, all HookNet models perform comparably in fatty tissue and other tissue classes, as can be observed in Figure \ref{fig:breast_performance}.

{\color{black} In lung tissue segmentation, the best HookNet (with $\lambda$=1.0) outperforms U-net(0.5) on the classes TLS and GC with an increase of 0.03 and 0.1 in $F_1$ score, respectively, and at the same time shows a decrease in $F_1$ score for Tumor by 0.01. $F_1$ scores for the 'other' class are the same for both models. A mixture of different models are outperforming HookNet on all distinct classes (U-Net(1.0) for TLS, U-Net(2.0) for GC, U-Net(0.5) for Tumor, and MRN for Other). However, HookNet achieves the highest overall $F_1$ score.}


{\color{black} We observed that HookNet, using the same fields of view as MRN, performs better than MRN, for both the breast and lung tissue segmentation, with respect to the overall $F_1$ score.
Finally, we observe that for breast tissue segmentation, HookNet(0.5, 8.0) performs best when giving more importance to the target branch (i.e., $\lambda$=75), while for the lung tissue segmentation the best $F_1$ scores are obtained when ignoring the context loss (i.e., $\lambda$ = 1).
}

{\color{black} To verify if there is a significant difference between HookNet and other models with the same target resolution (i.e., 0.5$\mu m/px$) we calculated the F1 score per test slide. We applied the Wilcoxon test, which revealed that for the breast dataset, the difference between HookNet and U-Net (p-value=0.004), and HookNet and MRN (p-value=0.001) are statistically significant. For the lung dataset, the differences between HookNet and U-Net (p-value=0.442), and HookNet and MRN (p-value=0.719) are not statistically significant. These results suggest that HookNet substantially benefits from wide contextual information (e.g., 8.0$\mu m/px$ for the input resolution), whereas the added value of context may be less prominent, but still beneficial, when relevant contextual information is restricted (e.g., 2.0$\mu m/px$ for the input resolution). Nonetheless, we argue that HookNet, based on the improvements made on TLS and GC, as can be seen by the F1 scores (see Table \ref{f1lung}) and in the confusion matrix (see Figure \ref{fig:confusion_matrices_lung}), can reduce the confusion between classes that are subjected to contextual information. }

\begin{table}[t]
\scriptsize
\input{f1unetsbreast}

\label{f1unetsbreast}

\bigskip

\input{f1breast}

\label{f1breast}

\end{table}

\begin{table}[t]
\input{f1unetslung}

\label{f1unetslung}

\bigskip

\input{f1lung}

\label{f1lung}
\end{table}

\newpage

\begin{figure*}[h]
\centering

\begin{subfigure}[b]{.18\linewidth}
\caption{\tiny Breast Tissue stained with H\&E}
\includegraphics[width=\linewidth]{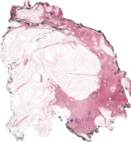}
\end{subfigure} \quad
\begin{subfigure}[b]{.18\linewidth}
\caption{\tiny U-Net ($0.5 \mu m/px$)}
\includegraphics[width=\linewidth]{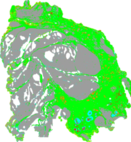}
\end{subfigure} \qquad
\begin{subfigure}[b]{.18\linewidth}
\caption{\tiny U-Net ($8.0 \mu m/px$)}
\includegraphics[width=\linewidth]{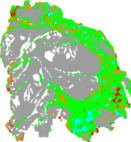}
\end{subfigure}\quad 
\begin{subfigure}[b]{.18\linewidth}
\caption{\tiny HookNet ($0.5, 8.0 \mu m/px$)}
\includegraphics[width=\linewidth]{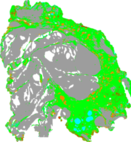}
\end{subfigure}

\vspace{3mm}

\begin{subfigure}{.18\linewidth}
\includegraphics[width=\linewidth]{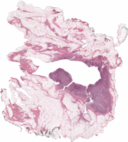}
\end{subfigure}\quad 
\begin{subfigure}{.18\linewidth}
\includegraphics[width=\linewidth]{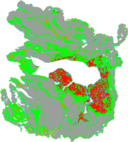}
\end{subfigure}\quad 
\begin{subfigure}{.18\linewidth}
\includegraphics[width=\linewidth]{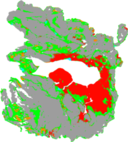}
\end{subfigure}\quad 
\begin{subfigure}{.18\linewidth}
\includegraphics[width=\linewidth]{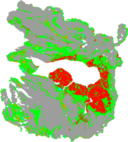}
\end{subfigure}

\vspace{3mm}

\begin{subfigure}{.18\linewidth}
\includegraphics[width=\linewidth]{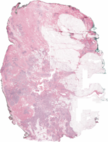}
\end{subfigure}\quad 
\begin{subfigure}{.18\linewidth}
\includegraphics[width=\linewidth]{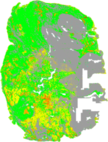}
\end{subfigure}\quad 
\begin{subfigure}{.18\linewidth}
\includegraphics[width=\linewidth]{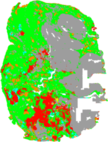}
\end{subfigure}\quad 
\begin{subfigure}{.18\linewidth}
\includegraphics[width=\linewidth]{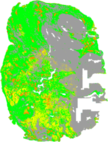}
\end{subfigure}

\vspace{3mm}

\begin{subfigure}{.05\linewidth}
\includegraphics[width=\linewidth]{images/performance_images/label_colors/dcis_label_color.png}
\end{subfigure}
\hspace{-3mm}
\begin{subfigure}{.10\linewidth}
\tiny DCIS
\end{subfigure}
\hspace{-2mm}
\begin{subfigure}{.05\linewidth}
\includegraphics[width=\linewidth]{images/performance_images/label_colors/idc_label_color.png}
\end{subfigure}
\hspace{-3mm}
\begin{subfigure}{.10\linewidth}
\tiny IDC
\end{subfigure}
\hspace{-2mm}
\begin{subfigure}{.05\linewidth}
\includegraphics[width=\linewidth]{images/performance_images/label_colors/ilc_label_color.png}
\end{subfigure}
\hspace{-3mm}
\begin{subfigure}{.10\linewidth}
\tiny ILC
\end{subfigure}
\hspace{-2mm}
\begin{subfigure}{.05\linewidth}
\includegraphics[width=\linewidth]{images/performance_images/label_colors/benign_label_color.png}
\end{subfigure}
\hspace{-3mm}
\begin{subfigure}{.10\linewidth}
\tiny Benign \\Epithelium
\end{subfigure}
\hspace{-2mm}
\begin{subfigure}{.05\linewidth}
\includegraphics[width=\linewidth]{images/performance_images/label_colors/normal_label_color.png}
\end{subfigure}
\hspace{-3mm}
\begin{subfigure}{.10\linewidth}
\tiny Other
\end{subfigure}
\hspace{-2mm}
\begin{subfigure}{.05\linewidth}
\includegraphics[width=\linewidth]{images/performance_images/label_colors/fat_label_color.png}
\end{subfigure}
\hspace{-3mm}
\begin{subfigure}{.10\linewidth}
\tiny Fat
\end{subfigure}




\caption{WSI predictions for the different models. First row: WSI example with DCIS. Second row: WSI example with IDC. Third row: WSI example with ILC.}
\label{fig:wsi_predictions}
\end{figure*}

\clearpage
\section{Discussion}

The main outcome of this research paper is two-fold.
%
The first outcome is a framework to effectively combine information from context and details in histopathology images.
We have shown its effect in segmentation tasks, in comparison with other single-resolution approaches, and with one multi-resolution recently presented.
The presented framework takes MFMR patches as input, and applies a series of convolutional and pooling layers, ensuring that feature maps are combined according to (1) the same spatial resolution, without the needs for arbitrary up-scaling and interpolation, as done in \cite{gu_multi-resolution_2018}, but allowing a direct concatenation of feature maps from the context branch to the target branch; (2) pixel-wise alignment, effectively combined with the use of \emph{valid} convolutions, which mitigates the risk of artifacts in the output segmentation map.
The optimal combination of fields of view used in the two branches of HookNet has been determined experimentally.
We first tested single-resolution U-Net models and then combined the fields of view that showed the best performance in two critical aspects of the problem-specific segmentation task, namely segmentation of DCIS and ILC {\color{black} for the breast dataset, and Tumor and GC for the lung dataset.}
At the moment, no procedure exists to select the optimal combination of spatial resolutions a priori, and empirical case-based analysis is needed.

The second outcome consists of two models for multi-class semantic segmentation in breast and lung cancer histopathology samples stained with H\&E.
In both cases, we have included \emph{tumor} as one of the classes to segment, as well as other types of tissue that can be present in the tumor tissue compartment, and made a specific distinction between three breast cancer subtypes, namely DCIS, IDC, and ILC.
Although a specific set of classes in breast and lung cancer tissue samples have been used as applications to show the potential of HookNet, presented methods are general and extendable to an arbitrary number of classes, as well as applicable to histopathology images of other organs.
Qualitative examples of segmentation output at whole-slide image level are depicted in Figure \ref{fig:wsi_predictions}, which shows the potential for using the outcome of this paper in several applications.
Segmentation of TLS and GC in lung squamous cell carcinoma can be used to automate TLS detection in lung cancer histopathology images, which will allow us to easily scale the analysis to a large number of cases, with the aim of further investigating the prognostic and predictive value of TLS count.
At the same time, segmentation of tumor and other tissue types allows to describe the morphology and tissue architecture of the tumor microenvironment, for example identifying the region of the tumor bulk, or the interface between tumor and stroma, an active research topic in immune-oncology, due to the role of tumor-infiltrating lymphocytes (TILs), which have to be assessed in the tumor-associated stroma (\cite{salgado_evaluation_2015}) as well as in the tumor bulk and at the invasive margin (\cite{galon_type_2006}, \cite{galon_towards_2014}). 
Furthermore, segmentation of both benign and malignant epithelial cells in breast cancer can be used as the first step in an automated pipeline for breast cancer grading, where the tumor region has to be identified to perform \emph{mitotic count}, and regions of both healthy and cancer epithelial cells have to be compared to assess \emph{nuclear pleormophism}.

In order to show the advantage of a multi-resolution approach compared to a single-resolution model in semantic segmentation of histopathology images, several design choices have been made in this paper.
Our future research will be focused on investigating the general applicability and design of HookNet with respect to the used constraints.
First, U-Net was used as the base model for HookNet branches. This choice was motivated by the effectiveness and flexibility of the encoder-decoder U-Net model, as well as the presence of skip connections. Other encoder-decoder models can be adopted to build a HookNet model.
Second, inspired and motivated by the multi-resolution nature of WSIs, we developed and solely applied HookNet to histopathology images.
However, we argue that HookNet has the potential to be useful for any application where a combination of context and details is essential to produce an accurate segmentation map.
Several applications of HookNet can be found in medical imaging, but it has the potential for being extended to natural images as well.
Third, we showed that using two branches allows to take advantage of clear trends like the performance of single-resolution models in DCIS and ILC (see {\color{black}Figure \ref{f1unetsbreast}}) in breast cancer data.
However, when focusing on the IDC class, we note that a single-resolution U-Net performs best at intermediate resolutions. This motivates further research in incorporating more branches, to include intermediate fields of view as well.
Fourth, we limited HookNet, as well as models used in comparison to 50M parameters, which allow model training using a single modern GPU with 11GB of RAM.
Introducing more branches will likely require a multi-GPU approach, which would also allow for experimenting with deeper/wider networks, and will speed-up inference time.


We compared HookNet in a single-loss {\color{black}($\lambda$ = 1) and in a multi-loss setup ($\lambda$=0.75, 0.5, or 0.25). Our results showed that the multi-loss model, when giving more importance to the target branch (e.g., $\lambda$=0.75), performs best for the breast tissue segmentation, and that the single-loss model (e.g., $\lambda$=1.0) scores best for the lung tissue segmentation. 
Future work will focus on an extensive optimization search for the value of $\lambda$}.

Finally, we reported results in terms of $F_1$ scores for both the Radboudumc and TCGA datasets based on \emph{sparse} manual annotations.
Although this is a common approach to obtain a large heterogeneous set of data in medical imaging, we observed that this approach limits the assessment of performance in the transition zones of different tissue types.
Extending the evaluation to an additional set of densely annotated data is part of our future research as well as effort in generating such manual annotations.

\section{Conclusion}
In this paper, we proposed HookNet, a framework for high-resolution tissue segmentation. We applied the model to two different datasets, which all included high resolution and contextual dependent tissue. Our results show that the proposed model increases overall performance compared to single-resolution models and can simultaneously deal with both subtle differences at high resolution as well as contextual information.



\section*{Acknowledgments}
The authors would like to thank Sophie van den Broek and Frederike Haverkamp for their support in the process of making manual annotations of breast tissue.
The results shown in this paper are partly based upon data generated by the TCGA Research Network: https://www.cancer.gov/tcga. This project has received funding from the European Union's Horizon 2020 research and innovation programme under grant agreement No 825292 (ExaMode project, www.examode.eu); from the Junior Researcher grant from the Radboud Institute of Health Sciences (RIHS), Nijmegen, The Netherlands; and from the Alpe dHuZes / Dutch Cancer Society Fund, grant number KUN 2014-7032.

\bibliographystyle{model2-names.bst}\biboptions{authoryear}
\bibliography{references}

\end{document}

%% file: f1unetsbreast.tex
\caption{{\color{black}Performance of U-Net with different input resolutions on the Radboudumc test set of breast cancer tissue types. Performance are reported in terms of F1 score per tissue type: ductalcarcinoma in-situ (DCIS), invasive ductal carcinoma (IDC), invasive
lobular carcinoma (ILC) benign epithelium (BE), Other, and Fat, as well as overall score (Macro $F_1$) measured on all classes together.}}
\centering
\resizebox{0.5\textwidth}{!}{%
{\color{black}
\begin{tabular}{c|c|c|c|c|c|c|c|c}
\multicolumn{2}{c|}{\textbf{Models}} & \multicolumn{6}{c|}{\textbf{F1-score}}                             & \textbf{} \\ \cline{1-8}
model &
  \begin{tabular}[c]{@{}c@{}}resolution\end{tabular} &
  \textbf{DCIS} &
  \textbf{IDC} &
  \textbf{ILC} &
  \textbf{Benign} &
  \textbf{Other} &
  \textbf{Fat} &
  \textbf{Overall} \\ \hline
\textbf{U-Net}    & 0.5  & 0.47          & 0.55         & \textbf{0.85} & 0.75  & 0.95 & 0.99 & 0.76      \\ \hline
\textbf{U-Net}    & 1.0   & 0.67          & 0.69         & 0.79 & \textbf{0.87} & \textbf{0.98} & 1.00 & 0.83      \\ \hline
\textbf{U-Net}    & 2.0  & 0.79          & 0.83         & 0.79 & 0.84          & 0.98 & 1.00 & \textbf{0.87}      \\ \hline
\textbf{U-Net}    & 4.0 & 0.83          & \textbf{0.85} & 0.63 & 0.73          & 0.96 & 1.00 & 0.83      \\ \hline
\textbf{U-Net}    & 8.0  & \textbf{0.86} & 0.81         & 0.20 & 0.66          & 0.96 & \textbf{1.00} & 0.75      \\ \hline
\end{tabular}%
}
}


%% file: f1breast.tex
\caption{{\color{black}Performance of U-Net(0.5), Multi-Resolution Network (MRN) from  Gu et al.
(2018), and HookNet on the Radboudumc test set of breast cancer tissue types.
Performance are reported in terms of F1 score per tissue type: ductal
carcinoma in-situ (DCIS), invasive ductal carcinoma (IDC), invasive
lobular carcinoma (ILC) benign epithelium (BE), Other, and Fat, as
well as overall score (Macro $F_1$) measured on all classes together.}}
\centering
\resizebox{0.5\textwidth}{!}{%
{\color{black}
\begin{tabular}{c|c|c|c|c|c|c|c|c|c|c}
\multicolumn{4}{c|}{\textbf{Models}} & \multicolumn{6}{c|}{\textbf{F1-score}}                             & \textbf{} \\ \cline{1-10}
model &
  \begin{tabular}[c]{@{}c@{}}target\\ resolution\end{tabular} &
  \begin{tabular}[c]{@{}c@{}}context \\ resolution\end{tabular} &
  $\lambda$ &
  \textbf{DCIS} &
  \textbf{IDC} &
  \textbf{ILC} &
  \textbf{Benign} &
  \textbf{Other} &
  \textbf{Fat} &
  \textbf{Overall} \\ \hline\hline
\textbf{U-Net}    & 0.5 & N/A & N/A  & 0.47          & 0.55         & 0.85      & 0.75          & 0.95 & 0.99 & 0.76      \\ \hline
\textbf{MRN}      & 0.5 & 8.0 & N/A  & 0.72          & 0.75         & 0.81      & 0.74          & 0.92 & 1.00 & 0.83      \\ \hline\hline
\textbf{HookNet}  & 0.5 & 2.0 & 1.0  & 0.62          & 0.75         & 0.82      & 0.82          & 0.98 & 1.00 & 0.83      \\ \hline
\textbf{HookNet}  & 0.5 & 8.0 & 1.0  & 0.84          & \textbf{0.9} & 0.86      & 0.80          & 0.97 & 1.00 & 0.90      \\ \hline
\textbf{HookNet}  & 0.5 & 8.0 & 0.5  & 0.83          & 0.87         & 0.88      & 0.76          & 0.98 & 1.00 & 0.89      \\ \hline
\textbf{HookNet}  & 0.5 & 8.0 & 0.25 & 0.83          & 0.88         & 0.81      & 0.72          & 0.97 & 1.00 & 0.87      \\ \hline
\textbf{HookNet} & 0.5 &  8.0 & 0.75 & \textbf{0.84} & 0.89 &    \textbf{0.91} & \textbf{0.84} & \textbf{0.98} & \textbf{1.00} & \textbf{0.91}
\end{tabular}%
}
}

%% file: f1unetslung.tex

\caption{{\color{black}Performance of U-Net trained with different input resolutions on the TCGA test set of lung cancer tissue. Performance are reported in terms of F1 score per tissue type: tertiary lymphoid structures (TLS), germinal centers (GC), Tumor, and Other, as well as overall score (Macro $F_1$) measured on all classes together.}}
\centering
\label{tab:my-table}
\resizebox{0.4\textwidth}{!}{%
{\color{black}
\begin{tabular}{c|c|c|c|c|c|c}
\multicolumn{2}{c|}{\textbf{Models}} & \multicolumn{4}{c|}{\textbf{F1-score}}                         & \textbf{}     \\ \cline{1-6}
model &
  \begin{tabular}[c]{@{}c@{}}resolution\end{tabular} &
  \textbf{TLS} &
  \textbf{GC} &
  \textbf{Tumor} &
  \textbf{Other} &
  \textbf{Overall} \\ \hline
\textbf{U-Net}    & 0.5   & 0.81          & 0.38          & \textbf{0.75} & \textbf{0.87}      & 0.70          \\ \hline
\textbf{U-Net}    & 1.0   & \textbf{0.86} & 0.44          & 0.71          & 0.86          & \textbf{0.72}          \\ \hline
\textbf{U-Net}    & 2.0  & 0.84          & \textbf{0.49} & 0.67          & 0.85          & 0.71          \\ \hline
\textbf{U-Net}    & 4.0   & 0.80          & 0.37          & 0.56          & 0.80          & 0.63          \\ \hline
\textbf{U-Net}    & 8.0   & 0.78          & 0.35          & 0.39          & 0.77          & 0.57          \\ \hline
\end{tabular}%
}
}



%% file: f1lung.tex

\caption{{\color{black}Performance of U-Net(0.5), Multi-Resolution Network (MRN) from  Gu et al. (2018), and HookNet  on the TCGA test set of lung cancer tissue. Performance are reported in terms of F1 score per tissue type: tertiary lymphoid structures (TLS), germinal centers (GC), Tumor, and Other, as well as overall score (Macro $F_1$) measured on all classes together.}}
\centering
\label{tab:my-table}
\resizebox{0.4\textwidth}{!}{%
{\color{black}
\begin{tabular}{c|c|c|c|c|c|c|c|c}
\multicolumn{4}{c|}{\textbf{Models}} & \multicolumn{4}{c|}{\textbf{F1-score}}                         & \textbf{}     \\ \cline{1-8}
model &
  \begin{tabular}[c]{@{}c@{}}target\\ resolution\end{tabular} &
  \begin{tabular}[c]{@{}c@{}}context \\ resolution\end{tabular} &
  $\lambda$ &
  \textbf{TLS} &
  \textbf{GC} &
  \textbf{Tumor} &
  \textbf{Other} &
  \textbf{Overall} \\ \hline\hline
\textbf{U-Net}    & 0.5 & N/A & N/A  & 0.81          & 0.38          & \textbf{0.75} & 0.87          & 0.70          \\ \hline
\textbf{MRN}      & 0.5 & 2.0 & N/A  & 0.83          & 0.40          & 0.71          & \textbf{0.88} & 0.71          \\ \hline\hline
\textbf{HookNet}  & 0.5 & 2.0 & 1.0  & 0.84          & \textbf{0.48} & 0.72          & 0.87          & \textbf{0.73} \\ \hline
\textbf{HookNet}  & 0.5 & 2.0 & 0.5  & 0.85          & 0.48          & 0.68          & 0.85          & 0.72          \\ \hline
\textbf{HookNet}  & 0.5 & 2.0 & 0.25 & 0.82          & 0.43          & 0.69          & 0.86          & 0.70          \\ \hline
\textbf{HookNet}  & 0.5 & 2.0 & 0.75 & \textbf{0.85} & 0.45          & 0.71          & 0.87          & 0.72         
\end{tabular}%
}
}